\documentclass[12pt,longbibliography]{article}

\usepackage[utf8]{inputenc}
\usepackage[resetlabels,labeled]{multibib}
\newcites{New}{Supplementary - References}
\usepackage{palatino}
\usepackage[T1]{fontenc}

\usepackage[a4paper, margin=2.5cm]{geometry}

\usepackage{amsmath, amsthm, amsfonts, amssymb}
\usepackage{mathrsfs} 
\usepackage{multirow}
\usepackage{xcolor}
\usepackage{subcaption}
\usepackage{graphicx}
\usepackage[colorlinks=true,linkcolor=blue,citecolor=blue,urlcolor=blue,breaklinks]{hyperref}

\usepackage{bbm}

\usepackage{breakurl}
\usepackage{natbib}
\usepackage{graphicx}

\theoremstyle{definition}

\theoremstyle{remark}

\theoremstyle{example}

\title{\large \textbf{Disentangling Community-level Changes in Crime Trends During the COVID-19 Pandemic in Chicago \\ \color{blue} \normalsize Published in \textit{Crime Science}: \href{https://crimesciencejournal.biomedcentral.com/articles/10.1186/s40163-020-00131-8}{https://crimesciencejournal.biomedcentral.com/articles/10.1186/s40163-020-00131-8\footnote{Please cite: Campedelli, G. M., Favarin, S., Aziani, A., \& Piquero, A. R. (2020). Disentangling community-level changes in crime trends during the COVID-19 pandemic in Chicago. \textit{Crime Science, 9}(1), 1-18. DOI: \href{https://doi.org/10.1186/s40163-020-00131-8}{https://doi.org/10.1186/s40163-020-00131-8}}}}}
\author{\normalsize Gian Maria Campedelli$^{\ddagger, 1}$ \and \normalsize  Serena Favarin$^{2,3}$ \and \normalsize  Alberto Aziani$^{2,3}$ \and \normalsize  Alex R. Piquero$^{4,5}$ }

\date{\footnotesize{$^{\ddagger}$ Corresponding Author: \texttt{gianmaria.campedelli@unitn.it}\\
$^1$ Department of Sociology and Social Research - University of Trento, Trento (Italy)\\
$^2$ Department of Political and Social Sciences - Università Cattolica del Sacro Cuore, Milan (Italy) \\
$^3$ Transcrime - Joint Research Centre on Transnational Crime, Milan (Italy)} \\
$^4$ University of Miami, Coral Gables, Florida (United States of America) \\
$^5$ Monash University, Melbourne, Victoria (Australia)\\
}

\begin{document}
\newgeometry{top=0.5cm,bottom=2cm,right=2cm,left=2cm}
\maketitle

\begin{abstract}
Recent studies exploiting city-level time series have shown that, around the world, several crimes declined after COVID-19 containment policies have been put in place. Using data at the community-level in Chicago, this work aims to advance our understanding on how public interventions affected criminal activities at a finer spatial scale. The analysis relies on a two-step methodology. First, it estimates the community-wise causal impact of social distancing and shelter-in-place policies adopted in Chicago via Structural Bayesian Time-Series across four crime categories (i.e., burglary, assault, narcotics-related offenses, and robbery). Once the models detected the direction, magnitude and significance of the trend changes, Firth’s Logistic Regression is used to investigate the factors associated to the statistically significant crime reduction found in the first step of the analyses. Statistical results first show that changes in crime trends differ across communities and crime types. This suggests that beyond the results of aggregate models lies a complex picture characterized by diverging patterns. Second, regression models provide mixed findings regarding the correlates associated with significant crime reduction: several relations have opposite directions across crimes with population being the only factor that is stably and positively associated with significant crime reduction.

%\normalsize
%\hspace{-2pt}\vspace{1\baselineskip}

\end{abstract}

%\tableofcontents
\normalsize
\restoregeometry

\section{Introduction}
The onset and spread of COVID-19 has affected nearly every continent and many millions of people. Not only has the virus infected, sickened, and killed scores of individuals as the virus moved from Asia, to Europe, to the US, and onwards to Central and South America, but it has also affected the lives of persons living in each of the countries that have experienced widespread infection. In particular, at the behest of public health officials, citizens of just about every country have been subject to social distancing measures, stay-at-home policies, shelter-in-place designations, and many were forced to remain in their homes for well over a month at a time in order to contain the spread of the virus.

As a result of these policy prescriptions, a wide array of research questions can be considered. For present purposes, we focus on the extent to which various policies associated with containing the spread of COVID-19 may have affected the frequency and patterning of criminal activity. Our interest in this space is not novel, as many have started to examine these questions. At the same time, most studies that do exist have used aggregate temporal series of crime data. These studies are certainly useful as they help to examine changes in crime before and after policies have been put into place at large units of analyses \citep{AshbyInitialevidencerelationship2020, AbramsCOVIDCrimeEarly2020, CampedelliExploringEffect2019nCoV2020, LeslieShelteringPlaceDomestic2020, MohlerImpactsocialdistancing2020, PiqueroStayingHomeStaying2020, PayneCOVID19SocialDistancing2020, PayneCOVID19ViolentCrime2020, HalfordCoronaviruscrimeSocial2020, GerellMinorcovid19association2020}. However, these studies have been unable to separate the spatial meso characteristics of changes in crime trends. The current study attends precisely to this issue. Herein, we use data from the 77 communities of Chicago to examine whether daily trends differ across crimes and communities and which factors are correlated with reductions in crime. Before we present the results of our study, we provide some background material in order to set the stage for our work and our specific analyses.
%\cite{koon,oreg,khar,zvai,xjon,schn,pond,smith,marg,hunn,advi,koha,mouse}

\section*{Background}

\subsection*{COVID-19 in Illinois and Chicago}
In the matter of a few months, the COVID-19 pandemic massively hit most regions of the world. Aside from its deleterious effects throughout the world, the United States is also experiencing the dramatic diffusion of the virus, currently being the country with the highest number of confirmed cases and deaths due to the disease. At the end of August, Illinois has the \color{black}sixth-highest number of confirmed cases of COVID-19 countrywide (more than 223,000 cases), with a death toll of almost 8,000 individuals. Cook County alone, which includes the city of Chicago and accounts for the highest portion of registered infections, reported more than 121,000 confirmed cases–i.e., about 2.3\% of the total population–with more than 5,000 deaths–i.e., about 4.15\% of lethality rate  \color{black} \citep{IllinoisInstituteofPublicHealthCoronavirusDisease20192020}.\color{black} 

Chicago represents the study site for the present investigation.
The diffusion of COVID-19 through February and early March of 2020, forced Illinois Governor Pritzker to first announce a disaster proclamation (i.e., a state of emergency) and subsequently issue official restrictive interventions to contain the diffusion of the virus. The first measures, such as school closures, were promulgated on March 13th, but it is with the set of interventions on March 15th that the State government started a series of escalating measures aimed at enforcing social distancing between citizens, such as the closure of bars and restaurants and the cancellation of all gatherings involving 50 or more people. On March 20th, the State government also issued a stay-at-home-order, closing all non-essential businesses. The stay-at-home order, which originally had to be in place until April 7th, was firstly extended for the entire State of Illinois until April 30th and, secondly, through May 29th.

\subsection*{Related Work}
The global COVID-19 emergency has caused unprecedented changes to interpersonal interactions and individual routine activities, and also to collective phenomena such as economies and political agendas. Crime is within the wide set of social processes and phenomena that COVID-19 has affected. In light of this, criminologists have started to explore whether and how the virus—or, more correctly, the virus together with the policies issued by governments—have changed the frequency of criminal activity. 
Works mainly analyzed trends across US cities, exploiting a variety of different quantitative methodologies \citep{MohlerImpactsocialdistancing2020, CampedelliExploringEffect2019nCoV2020, AshbyInitialevidencerelationship2020, LeslieShelteringPlaceDomestic2020, PiqueroStayingHomeStaying2020}. Statistical results were heterogeneous across contexts, with certain crimes exhibiting declines to different extents, while other remaining unvaried. Early works also studied the consequences of the policies in Australia \citep{PayneCOVID19SocialDistancing2020, PayneCOVID19ViolentCrime2020, PaynePropertyCrimeCOVID192020} and Europe, specifically in the United Kingdom \citep{HalfordCoronaviruscrimeSocial2020} and Sweden \citep{GerellMinorcovid19association2020}. 

While the research and policy importance of these studies is critical to address the newly emerging issues caused by the pandemic, further efforts are required to understand the meso-level patterns behind changes in crime trends. Decades of research in crime and delinquency have demonstrated that crime does not occur randomly across space and time \citep{FreemanSpatialConcentrationCrime1996b, Johnsonbriefhistoryanalysis2010, WeisburdLawCrimeConcentration2015a}. Events are rather clustered and follow specific patterns based on social, economic, demographic and ecological characteristics of the environment \citep{MertonSocialStructureAnomie1938, ShawJuveniledelinquencyurban1942, GibbsCrimeRatesAmerican1976, BlauCostInequalityMetropolitan1982, SampsonViolentvictimizationoffending1994, PapachristosMoreCoffeeLess2011, DammDoesGrowingHigh2014, WeisburdLawCrimeConcentration2015a}. The existence of such differences opens new lines of inquiry in the attempt to frame the impact that the COVID-19 pandemic has had on criminal phenomena. Accordingly, the current study attempts to disentangle meso-level dynamics focusing on Chicago’s communities, seeking to test whether changes in crime trends are consistent and similar across very different areas and, eventually, explore the main correlates associated with the actual presence of statistically significant reduction in daily-based crime counts. 

\subsection*{Theoretical Framework}
In an attempt to contribute both to the knowledge on the consequences of COVID-19 on crime and on the study of spatial and temporal patterns of crime in Chicago, we frame our work within the theoretical premises of routine activity theory \citep{CohenSocialChangeCrime1979}, crime pattern theory \citep{BrantinghamPatternsCrime1984} and general strain theory \citep{AgnewFoundationGeneralStrain1992}, while motivating our analytical framework from the perspective of the literature on crime concentration. 

Routine Activity Theory and Crime Pattern Theory outline that interactions and features of individual-level activities guide the spatio-temporal clustering of offending and victimization. Scholars have thus been interested in scrutinizing whether the modifications in human daily habits generated by the spread of COVID-19 have resulted in consequent changes in criminal activity. While Routine Activity Theory and Crime Pattern Theory point in the direction of a general crime reduction, especially for thefts, robberies, burglaries, and homicides, General Strain Theory goes in a different direction. In a situation of limited freedom of movement, increased social isolation, financial distress, and uncertainty related to the aforementioned containment policies and the risk of economic backlashes, individuals may be subjected to negative stimuli (e.g., stress) that could lead them to experience a range of negative emotions which, left unchecked, may lead to the commission of a crime. In light of this, while interaction and movement reduction can affect crime in the short-term, the prolonged stay-at-home order may trigger spikes in certain types of crimes in the medium- and long-ranges. 

Although the pandemic context represents an unprecedented scenario, we hypothesize that crime changes are not equally distributed across communities. The potential differences that one could observe within a city–a complex system made up of several social, economic, environmental layers–are thus tightly related to the literature on crime concentration \citep{ShawJuveniledelinquencyurban1942, FreemanSpatialConcentrationCrime1996b, Johnsonbriefhistoryanalysis2010}. For this reason, while Routine Activity Theory, Crime Pattern Theory and General Strain Theory are critical in picturing aggregate macro-dynamics, the literature on spatial concentration of crime opens further lines of inquiry that are crucial for better understanding the meso-level trends occurring in Chicago.

\section*{The Present Work}
\subsection*{Geographical and Criminal Focus}
The present work focuses on the 77 communities of Chicago. The aim of the study is twofold. First, analyze whether there exist community-level differences across crime trends in the post-COVID-19 containment policies period. Second, investigate statistical associations between significant crime reductions and four sets of correlates. These correlates map, respectively, (1) crime-related characteristics of each community, (2) socio-economic conditions, (3) health-related and demographic information and, finally, (4) the presence of joint pairwise reductions across crimes.

\color{black}The reasons for concentrating on crime reductions, rather than changes, are mainly two. First, previous studies highlighted that, with few important exceptions--e.g., domestic violence \citep{PiqueroStayingHomeStaying2020}--crime reduction is the prevalent spillover on crime of COVID-19 containment policies \citep{AshbyInitialevidencerelationship2020,CampedelliExploringEffect2019nCoV2020, MohlerImpactsocialdistancing2020,PayneCOVID19SocialDistancing2020}. The current study aims to better investigate this relevant dynamic. As second, the--likely--low numerosity of communities registering increases in crime counts, together with the relatively low number of units of analysis, may cause increases to bias any estimate.\color{black}

Chicago Community areas were created by the Social Science Research Committee at the University of Chicago in the 1920s and almost never changed since their original creation, except for O’Hare annexation to the city in 1956 and the separation of Edgewater from Upton in 1980. The division of Chicago in 77 communities mainly serves to organize service delivery and propose planning strategies and policies within the city. Each community area may comprise one or more neighborhoods. While other geographical divisions exist, the clear distinction of the boundaries across communities and their stable presence in time facilitated the process for which most of Chicago statistics are produced at the community-level.

Our analysis will focus on four crime types: burglaries, assaults, narcotics-related offenses, and robberies. Three sets of reasons lie behind the choice of these four crime categories. Firstly, these crimes (with the exception of narcotics) have been analyzed by recent studies focusing on the impact of COVID-19 on crime. Our study would thus provide further insights on crime typologies that have been investigated only at aggregate level. Second, these crime categories are distinct in nature, hence allowing us to observe not only differences at spatio-temporal levels but also crime-wise. Burglaries, for instance, are appropriative and are not mandatorily “physical”, per se. Contrarily, assaults, robberies and most of narcotics-related offenses require interaction between individuals. Furthermore, robberies and assaults are mostly characterized by a certain degree of violence. Also, while burglaries and robberies map appropriative offenses, assaults have a rather expressive nature and are generally motivated by different objectives or reasons. \color{black}Finally, narcotics-related offenses are connected to the provision of an illegal service, that can be regulated by complex criminal markets. At the same time, narcotics are also associated with policing activities. Thanks to these characteristics, narcotics allow for analyzing additional dynamics that possibly emerge as consequence of the anti-COVID-19 measures. \color{black}Thirdly, these four crime categories account for a total of more than 122,000 reported crimes registered in the Chicago Crime Database from January 1st 2018 to May 17th 2020,\footnote{The sum of the four crime categories accounts for more than 20\% of all registered crimes in the original dataset.} thus ensuring a sufficient number of crimes to carry out reliable analyses.

\subsection*{Methods}\label{methods}
This work relies on a two-step methodology. In the first part, we evaluate the impact of social distancing in Chicago, specifically focusing on burglaries, assaults, narcotics-related crimes, robberies across all of the 77 urban communities that are part of the city through the use of Bayesian Structural Time Series \citep{BrodersenInferringcausalimpact2015}, partially replicating the methodology presented in \cite{CampedelliExploringEffect2019nCoV2020}. Within the context of the Bayesian models, we focus on the distribution, direction, and significance of the relative cumulative effect (RCE) of COVID-19 containment policies for each crime in each community. The RCE captures the relative difference (rather than the crude count) between the actual number of crimes that occurred in the post-intervention period (which goes from March 16th to May 17th) and the predicted number of crimes that we would have expected in absence of any intervention, as computed by the simulation model. This first part of the methodology allows for understanding whether changes in crime trends (either positive or negative) exist and what their distribution across Chicago communities is. \color{black} The models are coded in R \citep{RCoreTeamLanguageEnvironmentStatistical2013} and performed using the CausalImpact package \citep{BrodersenInferringcausalimpact2015}. \color{black}

In the second part, we shift the object of our analysis on the detected statistically significant crime reductions. \color{black} For this reason, we have created a variable--“Crime Reduction” (CR)-mapping the actual presence of a statistically significant crime reduction. CR is equal to 1 if the relative cumulative effect is lower than 0 and the associated p-value is lower than 0.05 and takes 0 values otherwise: \color{black}

\[\mathrm{CR}=:\left\{\begin{matrix} 1 & \mathrm{if} & \left [\mathrm{RCE}<0 \right ] \wedge \left [p-val(\mathrm{RCE})\leq 0.05 \right ]\\ 0 & & \mathrm{otherwise} \end{matrix}\right.\]

Once the variable has been calculated for each crime in each community, four sets of regression models are performed to assess the correlates of observed crime trends at the community level. Firth’s Logistic Regression \citep{FirthBiasReductionMaximum1993} is employed as it is the most adapt statistical strategy to deal with particularly small samples (77 observations, namely the Chicago communities), and unbalanced classes--i.e., in our setup, an unequal number of communities in which crime trends significantly decreased \citep{Heinzesolutionproblemseparation2002, NemesBiasoddsratios2009}. Firth’s regression aims at reducing the bias that generally arises when using ordinary logistic regression with small and well-separated samples and specifically uses penalized likelihood instead of maximum likelihood estimation. 

\color{black}The choice to estimate separate regression models is motivated by the small size of the sample. The limited number of observations does not make it possible to estimate models with a large number of covariates, and we have thus decided to perform distinct theoretically-driven analyses to explore the effects of the selected predictors within each dimension of interest (e.g., the effect of poverty within the socio-economic dimension). Such approach, although forced by the structural nature of the sample, comes with limitations. First, it raises the likelihood of omitting relevant information and it raises the risk of obtaining spurious relationships. Second, it makes it unfeasile to compare effect size across models. While we acknowledge that estimating separate reduced models represents an inherent limitation of the study, we have tried to limit the risk of obtaining biased results by theoretically motivating the inclusion of our predictors and by being parsimonious to avoid overfitting. Furthermore,  not only we have used 2019 rates of burglaries, assaults, narcotics and robberies as main predictors in the crime-related models: we have also included them as controls in the other three dimensions. Including them as controls, we provide stronger evidence of the potential relation between our dependent variables and the crime-unrelated covariates chosen in the models. \color{black} The regression models are performed in Stata  \citep{StataCorpStataStatisticalSoftware2015} using the FIRTHLOGIT \citep{CoveneyFIRTHLOGITStatamodule2015} package.

\color{black}
\subsection*{Data}
\begin{figure}[hbt!]
        \centering
        \begin{subfigure}[b]{0.4\textwidth}
            \centering
            \includegraphics[width=\textwidth]{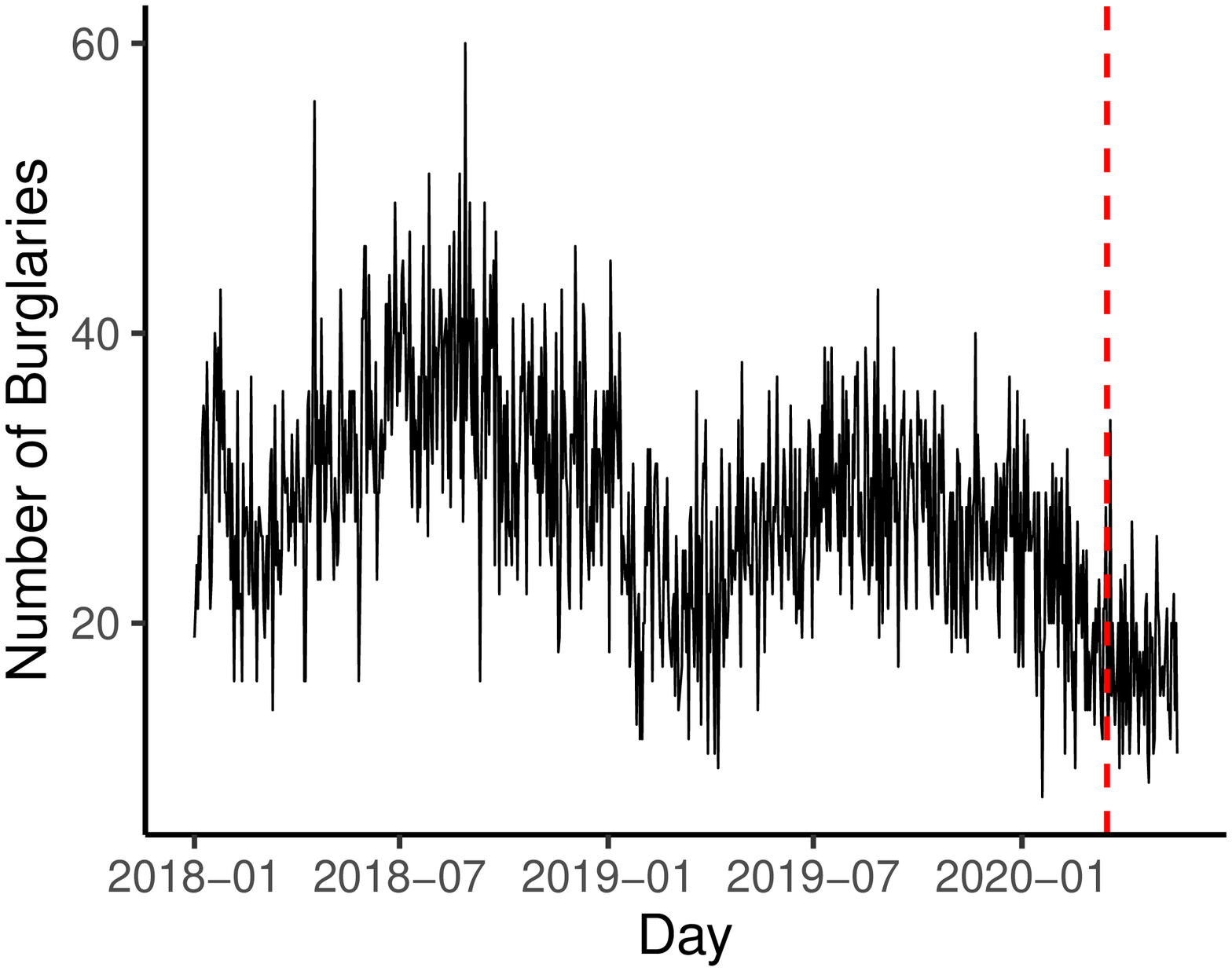}
            \caption[Network2]%
            {{\small Burglaries}}    
            \label{fig:mean and std of net14}
        \end{subfigure}
        %\hfill
        \begin{subfigure}[b]{0.4\textwidth}  
            \centering 
            \includegraphics[width=\textwidth]{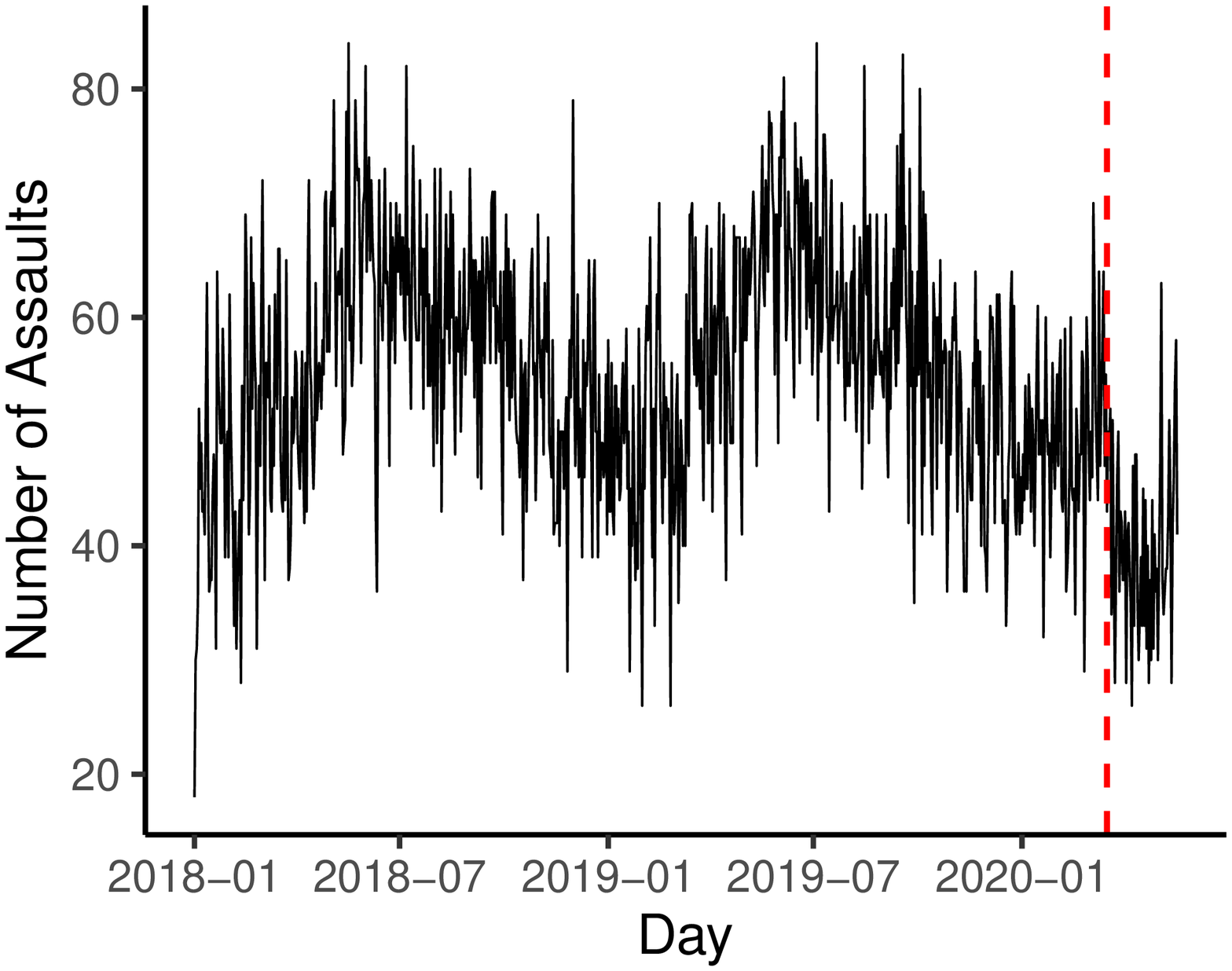}
            \caption[]%
            {{\small Assaults}}    
            \label{fig:mean and std of net24}
        \end{subfigure}
        \vskip\baselineskip
        \begin{subfigure}[b]{0.4\textwidth}   
            \centering 
            \includegraphics[width=\textwidth]{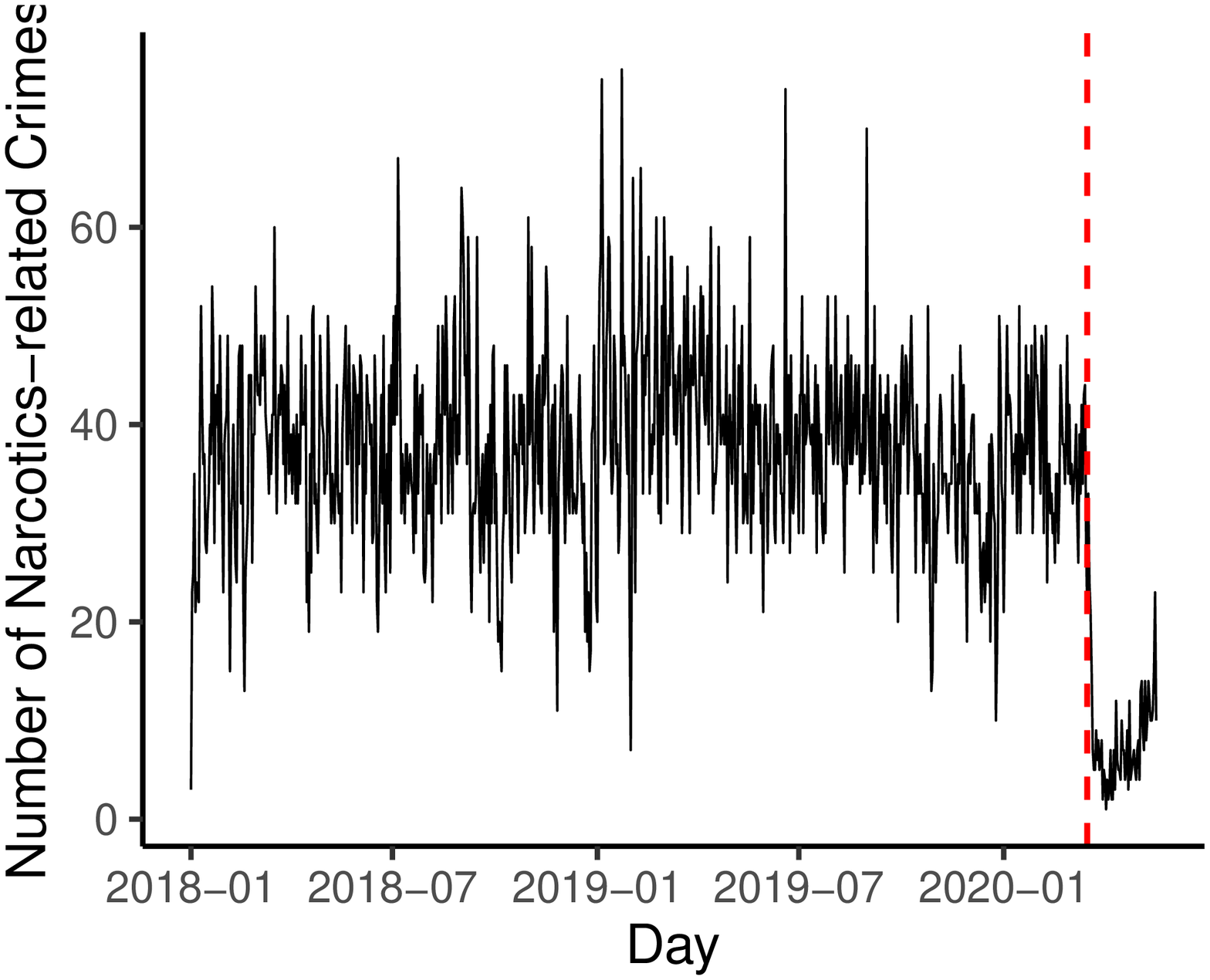}
            \caption[]%
            {{\small Narcotics}}    
            \label{fig:mean and std of net34}
        \end{subfigure}
        \quad
        \begin{subfigure}[b]{0.4\textwidth}   
            \centering 
            \includegraphics[width=\textwidth]{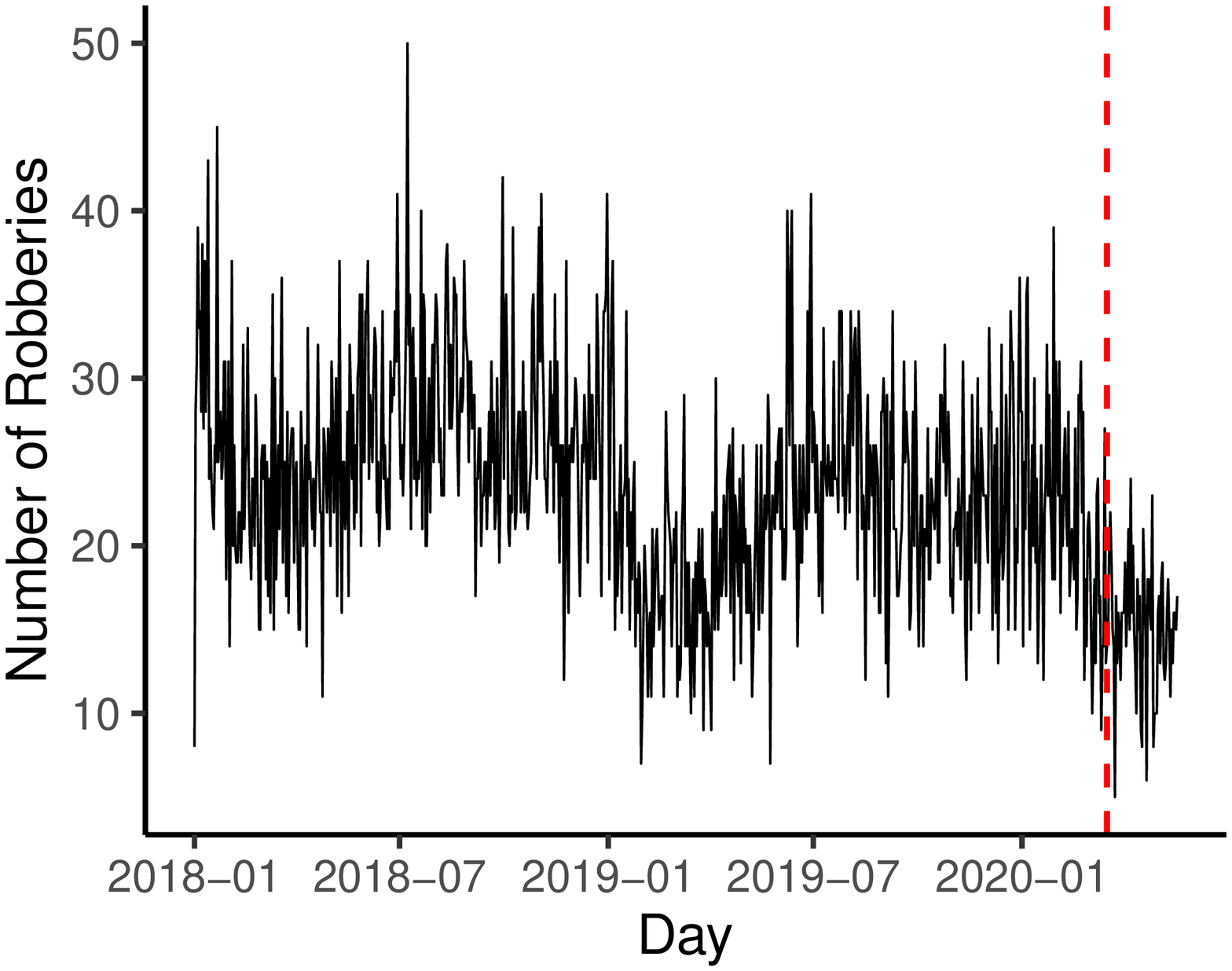}
            \caption[]%
            {{\small Robberies}}    
            \label{fig:mean and std of net44}
        \end{subfigure}
        \caption[\centering Daily counts per each crime (Entire City) - Dashed vertical line indicates date when containment policies began in Chicago (March 16th 2020)]
        {\small Daily counts per each crime (Entire City) - Dashed vertical line indicates date when containment policies began in Chicago (March 16th 2020)} 
        \label{fig:timeseries}
    \end{figure}

Several data sources have been employed to conduct the analyses. First, the primary dataset is gathered from the Chicago Data Portal and originally includes all the reported crime incidents (except for murders) that occurred in the city of Chicago from January 1st 2001 to present. For the purposes of this article, a subset of crime types has been temporally selected, namely all the burglaries, assaults, narcotics-related offenses, and robberies that occurred from January 1st 2018 to May 17th 2020.\footnote{In our analyses we use the date of crime commission rather than the date of crime reporting because we wanted to capture when a crime occurred–rather then when it was reported–to better estimate the daily changes in crime trends. To avoid possible biases due to late reporting we collected the data on June 06th 2020–three weeks after May 17th 2020 which is the last day included in our timeframe.} The decision to rely on such a time-frame has been made to help ensure that the models capture long-term dynamics, including monthly and quarter seasonality. In fact, analyses focusing on overly short time-frames face the risk of missing the presence of already existing trends originating in the medium- or long-range past. The extraction from the Chicago Data Portal led to a total of 122,421 offenses (descriptive statistics by community are reported in Table \ref{table:desc} and aggregated daily trends are displayed in Figure \ref{fig:timeseries}).\footnote{Since the dataset is based on data recorded by the Chicago Police Department’s Citizen Law Enforcement Analysis and Reporting (CLEAR) System, the coding of each crime follows the Illinois Uniform Crime Reporting codes: each of the four mentioned crimes thus include a list of specific offenses. Burglary specifically IUCR codes 0610, 0650, 0620, 0630; assault includes 0560, 0520, 0558, 051A, 0530, 0554, 0545, 0555, 0550, 0553, 0557, 051B, 0552, 0551, 0556. Narcotics specifically includes 2022, 2027, 2093, 2024, 2028, 1811, 1812, 1821, 2014, 2034, 1822, 2092, 2017, 2026, 2013, 2021, 2023, 2090, 2031, 2170, 2091, 2016, 2012, 2018, 2110, 2020, 2025, 2070, 2011, 2015, 2029, 2032, 2033, 1840, 1850, 2160, 2050, 2094, 2019, 2030, 2095, 2040, 2010, 2080. Finally, robbery includes 0325, 0320, 0312, 0330, 031A, 033A, 0326, 0334, 0331, 0340, 0313, 031B, 0337, 033B. For a complete list of IUCR codes see  \href{https://data.cityofchicago.org/Public-Safety/Chicago-Police-Department-Illinois-Uniform-Crime-R/c7ck-438e/data}{https://data.cityofchicago.org/Public-Safety/Chicago-Police-Department-Illinois-Uniform-Crime-R/c7ck-438e/data} }  

\begin{table}[!hbt]
\centering
\color{black}
\caption{Descriptive Statistics of Considered Crimes (By Community)}
\begin{tabular}{lcccccc}
\hline
\textbf{Crime Category} & \textbf{Obs (Overall)} & \textbf{Min} & \textbf{Max} & \textbf{Mean} & \textbf{Median} & \textbf{St. Dev.} \\\hline
Burglary & 24,068 & 22 & 1,174 & 312.57 & 249 & 242.74 \\
Assault & 47,197 & 41 & 2,937 & 612.95 & 410 & 544.53 \\
Narcotics & 31,044 & 2 & 4,539 & 403.16 & 127 & 820.05 \\
Robbery & 20,216 & 2 & 1,645 & 262.54 & 180 & 277.20\\\hline
\end{tabular}
\label{table:desc}
\end{table}

In the second part of the analyses, the results obtained from the Bayesian models are used to understand what factors are associated with a significant reduction in terms of RCE, particularly analyzing CR as the target variable of interest. As anticipated, four different categories of correlates are considered: a crime-related dimension, a socio-economic dimension, a health-demographic dimension and a so-called joint reduction dimension. Each of these dimensions is populated by several variables that are hypothesized to be linked to the presence (or absence) of a statistically significant crime reduction in a given community. The considered variables, along with their sources, are listed in Table \ref{dimensions}.
\begin{table}[]
\footnotesize
\centering
\caption{Description of Dimensions, Variables and Sources}
\begin{tabular}{llll}
\hline
Dimension & Variable & Source & Time-Span \\\hline
\multirow{6}{*}{\begin{tabular}[c]{@{}l@{}}Crime-\\ related\end{tabular}} & Burglary   Rate (10k) & \cite{ChicagoPoliceDepartmentCrimes2001Present2020} & 2019 \\
 & Assault   Rate (10k) & \cite{ChicagoPoliceDepartmentCrimes2001Present2020} & 2019 \\
 & Narcotics   Rate (10k) & \cite{ChicagoPoliceDepartmentCrimes2001Present2020} & 2019 \\
 & Robbery   Rate (10k) & \cite{ChicagoPoliceDepartmentCrimes2001Present2020} & 2019 \\
 & Neighborhood   Safety & \cite{ChicagoDepartmentofPublicHealthChicagoNeighborhoodSafety2018} & 2016-2018 \\
 & Has   Police & \cite{ChicagoPoliceDepartmentPoliceStations2016} & 2016 \\\hline
\multirow{5}{*}{\begin{tabular}[c]{@{}l@{}}Socio-\\ Economic\end{tabular}} & Population/1000 & \cite{U.S.CensusChicagoTotalPopulation2016} & 2012-2016 \\
 & Income   Diversity & \cite{U.S.CensusChicagoIncomeDiversity2018} & 2014-2018 \\
 & Crowded Housing   Rate & \cite{U.S.CensusChicagoCrowdedHousing2015} & 2012-2016 \\
 & Vacant   Housing Rate & \cite{U.S.CensusChicagoVacantHousing2015} & 2012-2016 \\
 & Poverty   Rate & \cite{U.S.CensusChicagoPovertyRate2018} & 2018 \\\hline
\multirow{4}{*}{\begin{tabular}[c]{@{}l@{}}Health-\\ Demographic\end{tabular}} & \% People   Aged 65+ & \cite{HeartlandAllianceDataChicagoPeopleAged2018a} & 2018 \\
 & \% People   Aged \textless{}18 & \cite{HeartlandAllianceDataChicagoPeopleAged2018} & 2018 \\
 & Overall Health   Status & \cite{ChicagoDepartmentofPublicHealthChicagoOverallHealth2016} & 2016-2018 \\
 & COVID-19   Cases Rate & \citep{IllinoisNationalElectronicDiseaseSurveillanceSystemChicagoCOVID19Cases2020} & 2020 \\\hline
\multirow{4}{*}{\begin{tabular}[c]{@{}l@{}}Joint\\ Reduction\end{tabular}} & Burglary   Sig. Reduction & \cite{ChicagoPoliceDepartmentCrimes2001Present2020} & 2018-2020 \\
 & Assault   Sig. Reduction & \cite{ChicagoPoliceDepartmentCrimes2001Present2020} & 2018-2020 \\
 & Narcotics   Sig. Reduction & \cite{ChicagoPoliceDepartmentCrimes2001Present2020} & 2018-2020 \\
 & Robbery   Sig. Reduction & \cite{ChicagoPoliceDepartmentCrimes2001Present2020} & 2018-2020\\\hline
\end{tabular}
\label{dimensions}
\end{table}

\subsubsection*{Crime-related Context}
\color{black} The crime-related dimension includes data on burglary, assault, narcotics, and robbery rates in each community in reference to 2019. \color{black} The rationale behind the inclusion of these variables lies in the need for understanding whether significant crime reductions are driven by crime rates in the past. From a theoretical point of view, it is relevant to understand whether the most evident positive effects (i.e., crime reductions) are observed regardless of the pre-lockdown levels of criminal activity or if, contrarily, there exist patterns highlighting greater benefits experienced by areas generally known for their high crime presence. Furthermore, a measure mapping neighborhood safety is considered to investigate how individual self-perception is reflected in the criminal changes that occurred during the lockdown. \color{black} Neighborhood Safety is in fact a measure gathered from the Healthy Chicago Survey (provided by the Chicago Department of Public Health), which estimates the rate of adults aged 18 years and older who report to feel safe all of the time and most of the time in a given community. The rate is weighted to reproduce the population from which the sample is drawn. In line with our reasoning regarding crime rates, we are interesting in verifying whether communities that are perceived as less safe have benefited more from the COVID-19 containment policies.    \color{black} Finally, the dimension also controls for the presence of a police station within the community, as to contribute to the existing debate on the effects of police presence on crime trends \citep{WilsonEffectPoliceCrime1978, DiTellaPoliceReduceCrime2004, RatcliffeCrimeDiffusionDisplacement2011}. 

\subsubsection*{Socio-Economic Conditions}
Concerning the socio-economic dimension, five correlates are analyzed. First, the relationship between crime reduction and the number of inhabitants is investigated. The criminological and economic literature has widely investigated the variability in crime between urban and non-urban/rural contexts, and one of the characterizing elements of urban contexts is the higher number (and density) of inhabitants. Urbanization is positively associated to criminal activity \citep{FlangoPovertyUrbanizationCrime1976, LarsonCrimeJusticeSociety1996, GlaeserWhyThereMore1999}, but within the context of a pandemic-driven lockdown, the criminal opportunities related to the dynamic aspects of urban life (such as transient crowding, see \cite{JarrellTransientCrowdingCrime1990}) may be consistently reduced, thus thinning the role of urbanization–and population–in driving criminal activity. Furthermore, in line with routine activity theory \citep{CohenSocialChangeCrime1979}, highly populated areas may have developed a higher level of formal and informal guardianship and control, thus further minimizing criminal opportunities.

Secondly and thirdly, income diversity and poverty are considered. \color{black} Income inequality and poverty at the community level may be associated with the experience of status and monetary strain, which, in turn may contribute to generate anger and frustration \citep{AgnewGeneralStrainTheory1999}\color{black}. Research has gone beyond the borders of relative and absolute deprivation to investigate the interplay between the two concepts, finding mixed results that generally vary across units of analyses (e.g., countries, regions, cities, communities) and crime types \citep{PattersonPovertyIncomeInequality1991, HoogheUnemploymentInequalityPoverty2011, PareIncomeinequalitypoverty2014, BurrastonRelativeAbsoluteDeprivation2018}. In light of the complex relationship between income inequality and poverty, this study will test whether we can identify associations between the two and the significant reduction of crimes in Chicago communities.

A fourth factor regards the crowded housing rate. The forced condition of shelter-in-place calls for careful consideration of crowded housing rate as a potential additional risk factor for crime. A situation of forced cohabitation in poor social conditions where positively valued goals become difficult to achieve, and positive stimuli are inhibited while negative ones, such as stress and impoverishment, are introduced, fit well within the frame expressed by general strain theory \citep{AgnewFoundationGeneralStrain1992}. Finally, the vacant housing rate is also included in the models. Previous research has noted the role of abandoned buildings and places as attractors for crime \citep{VigilStreetsocializationLocura1988, SpelmanAbandonedbuildingsMagnets1993}: in the context of forced reduced human mobility and freedom to move, vacant houses and abandoned buildings may even acquire greater importance for criminal activity within a community. 

\subsubsection*{Health-related and Demographic Conditions}
Social scientists have long debated about the influence of age structure on crime, both in terms of temporal and geographical distributions \citep{CohenAgeStructureCrime1987, PhillipsRelationshipAgeStructure2006, ArnioDemographyforeclosurecrime2012}. Therefore, variables controlling for the percentage of people aged more than 65 years old and for the percentage of juveniles (people aged less than 18 years) are considered. Potential dynamics may arise in neighborhoods marked by higher levels of the elderly. With everyone forced in their residences and an overall increase in enforced guardianship, offenders may for instance target houses inhabited by seniors, hypothesizing a weaker type of guardianship and resistance. 

A third variable refers to the overall health status of the community. \color{black} The variable maps the estimated rate of adults (aged 18 and older) who reported their overall health status as being good, very good or excellent community-wise and the measure is weighted to reflect the structure of the population from which the sample is drawn. \color{black} Scientists in different disciplines have been interested in measuring how health status and fear of crime interact together \citep{BazarganEffectsHealthEnvironmental1994, McKeeHealthFearCrime2000, Chandolafearcrimearea2001}. Although actual and perceived crime-levels often differ \citep{WeatherburnCrimePerceptionReality1996, PfeifferMediaUseits2005, HippResidentPerceptionsCrime2010}, it is relevant to understand whether communities that, on average, perceived better status levels before the pandemic have been associated with a significant reduction of crimes. 

Finally, a variable expressing the rate of COVID-19 infected individuals living in the community serves the purpose of investigating whether areas with higher levels of contagion are also associated with higher (and significant) crime reduction patterns. A higher prevalence of cases may lead to higher levels of fear within the community. Fear of COVID-19, as empirically demonstrated \citep{HarperFunctionalFearPredicts2020}, has a positive role in strengthening compliant behaviors, including social distancing and personal hygiene, and may consequently indirectly reduce crime, by impacting daily actions of both non-offenders and offenders. \footnote{We preferred to use the rate of infected persons instead of the rate of dead persons, because the COVID-19 related deaths were often misidentified by the health care systems. Hospitals and local governments did not always test the people who died with COVID-19 symptoms, so we decided to use established number of infections in our analysis. Data on daily cases, deaths and hospitalizations are available here (information on addresses are included): https://data.cityofchicago.org/Health-Human-Services/COVID-19-Cases-Tests-and-Deaths-by-ZIP-Code/yhhz-zm2v/data.}

\subsubsection*{Other Crimes Evolution}
Finally, the fourth dimension tests whether a significant reduction in one crime is positively associated with a reduction in another one. Research on crime containment policies has demonstrated that, besides direct effects of interventions on criminal activity, a number of indirect consequences can possibly emerge, including displacement mechanics of various types \citep{ReppettoCrimePreventionDisplacement1976, GueretteAssessingExtentCrime2009, RatcliffeCrimeDiffusionDisplacement2011}. \color{black} Specifically, it might happen that reductions in one crime type in a certain area is followed by increases of other crimes in the same environment. Within the exceptional context of the lockdown, we may expect different crimes to follow common trends, due to a diffused reduction in criminal opportunities. Alternatively, we might observe a shift toward more expressive crimes induced by the simultaneous reduction of opportunities and increase in strain. \color{black}

\section*{Results}

\subsection*{Community-level crime-wise Bayesian models}
Overall, the first part of the analysis highlights that the effects of the COVID-19 containment policies are distinct in terms of geographic distribution, effect magnitude, significance, and prevalence across Chicago communities. The following maps show, for each crime, the presence of a significant change in the daily trend (top subfigures), along with the magnitude of the cumulative RCE (bottom subfigures).

\begin{figure}[hbt!]
\centering
\begin{subfigure}{0.45\textwidth}
  \centering
  \includegraphics[width=1\linewidth]{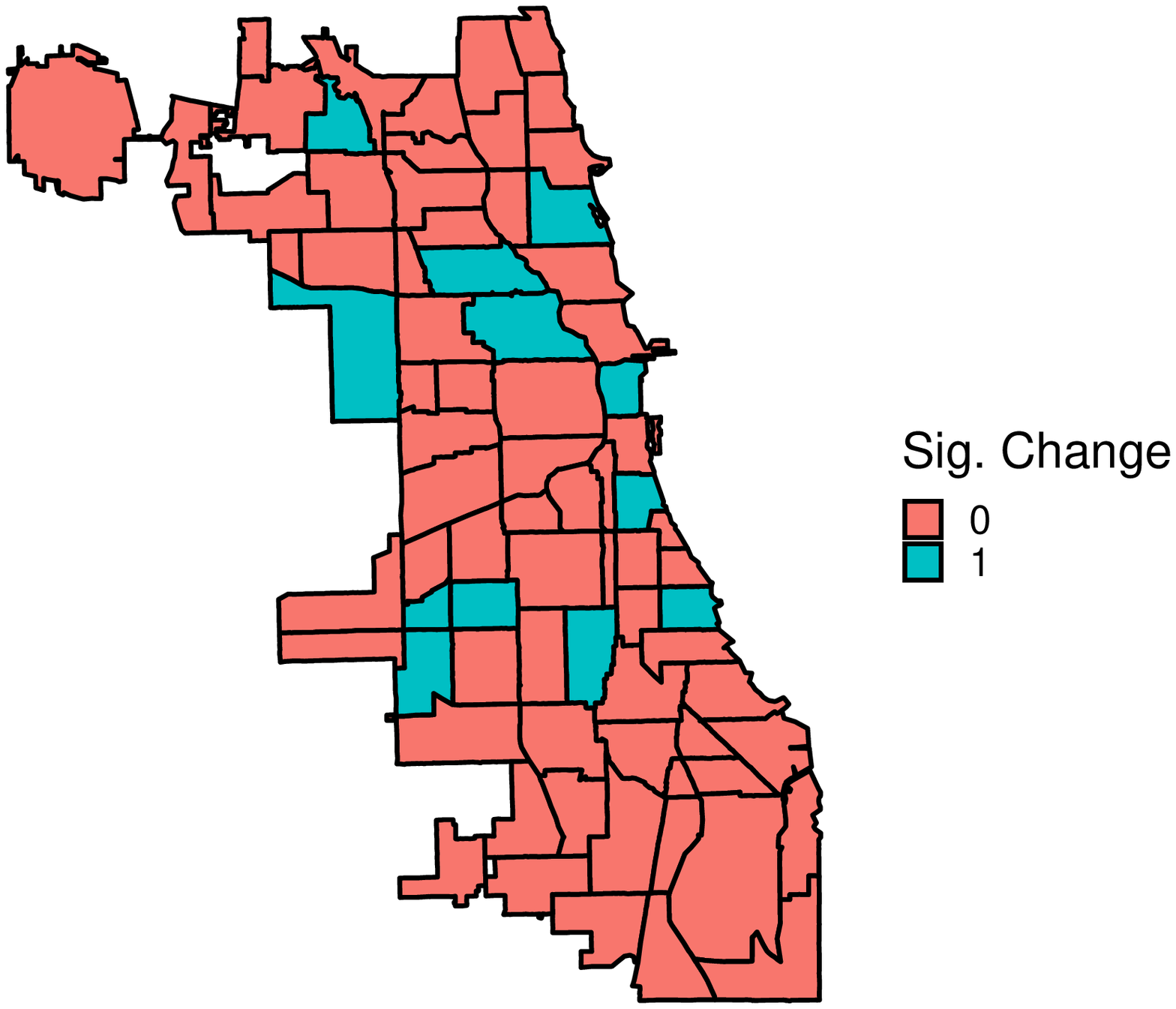}
  \caption{\footnotesize Communities Experiencing Statistically Significant Changes in Burglaries}
  \label{fig:sub1}
\end{subfigure}%
\begin{subfigure}{.45\textwidth}
  \centering
  \includegraphics[width=1\linewidth]{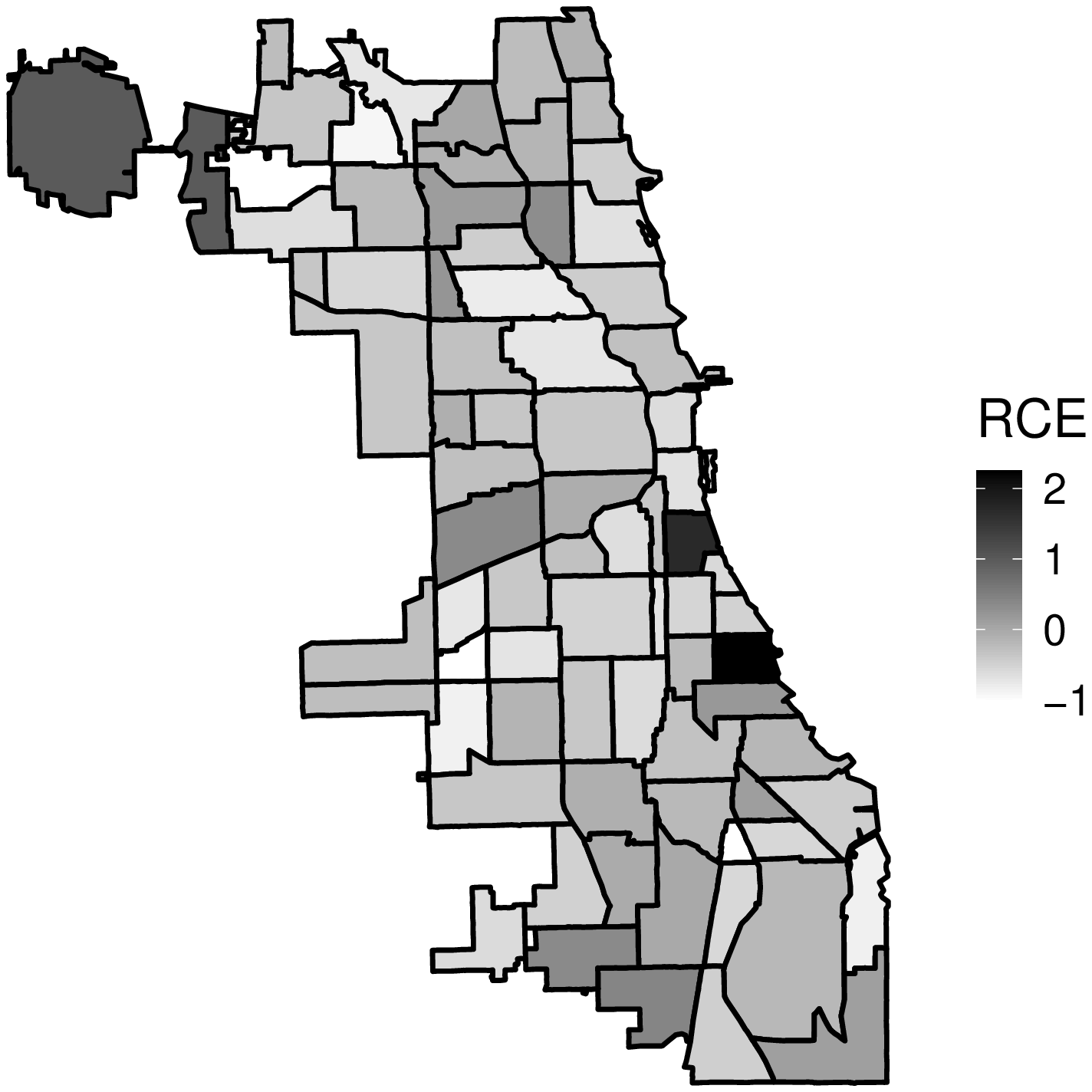}
  \caption{\footnotesize Burglaries RCE across Communities}
  \label{fig:sub2}
\end{subfigure}
\caption{}
\label{fig:b1}
\end{figure}

Figure \ref{fig:b1} refers to the community-wide analysis of burglaries. It is straightforward to notice that the vast majority of communities did not experience any significant change (i.e., red-colored communities). The bottom pictures the distribution of the RCE magnitude per each community (lighter colors are associated with stronger negative RCE values–i.e., crime reduction–while darker are linked to higher positive RCE-i.e., crime increase). The bottom map indicates that there is a relevant share of communities (n=12) that experienced an increase, even up to CRE=2.19 (i.e., an increase of 219\% compared to the \color{black} forecasted cumulative number of crimes as estimated by the counterfactual scenario with no issued interventions produced by the Bayesian model), although only 2 of these increases were found to be significant. The two communities that experienced a significant increase in crime are Douglas and Hyde Park, situated in the South Side district. Among the socio-structural factors that cause opposite trends in burglaries, different distributions of residential and non-residential buildings may play a role, as COVID-19 containment policies may impact residential and non-residential burglaries to different extents \citep{AshbyInitialevidencerelationship2020}.\color{black}

\begin{figure}[hbt!]
\centering
\begin{subfigure}{0.45\textwidth}
  \centering
  \includegraphics[width=1\linewidth]{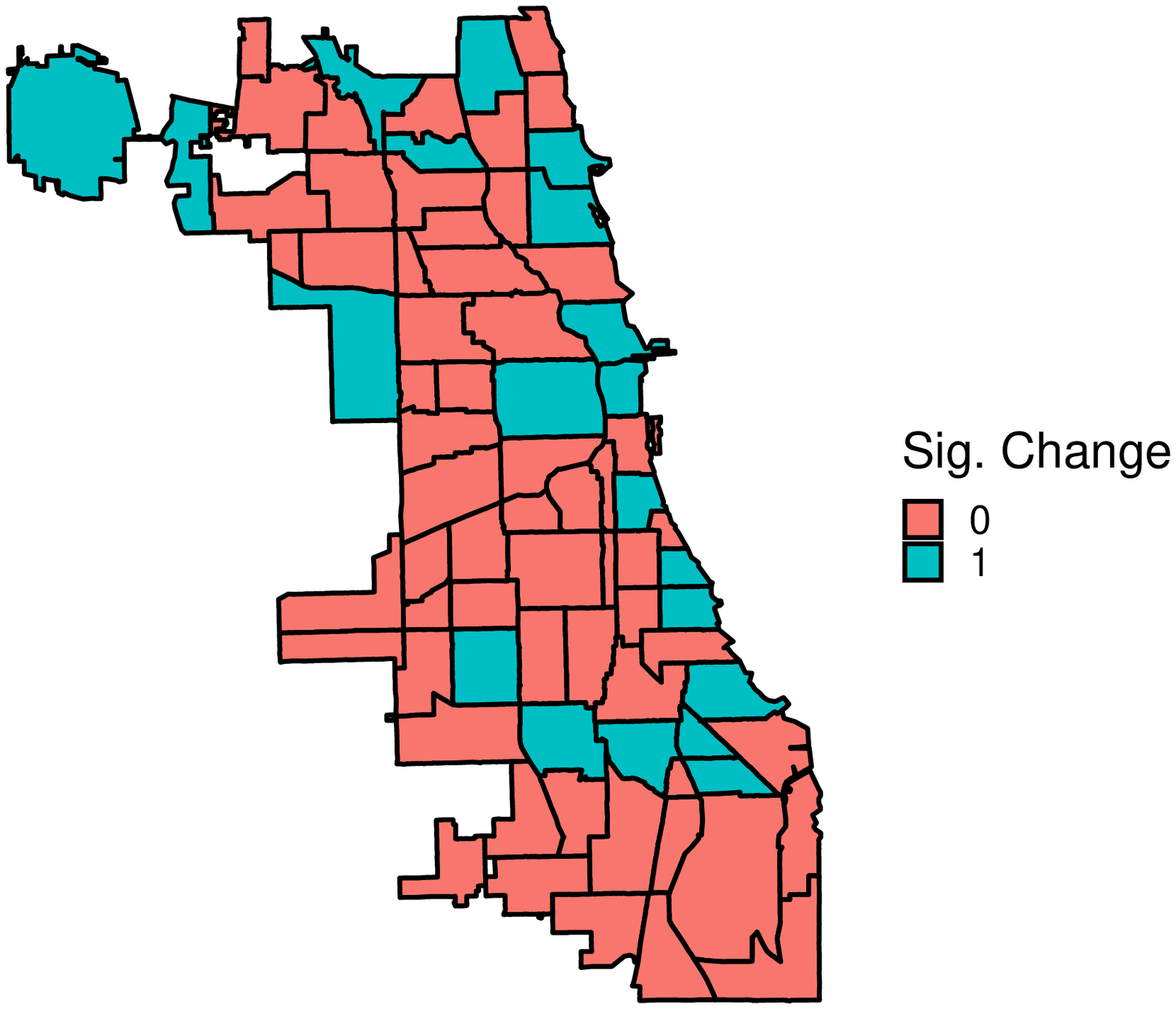}
  \caption{\footnotesize Communities Experiencing Statistically Significant Changes in Assaults}
  \label{fig:sub1}
\end{subfigure}%
\begin{subfigure}{.45\textwidth}
  \centering
  \includegraphics[width=1\linewidth]{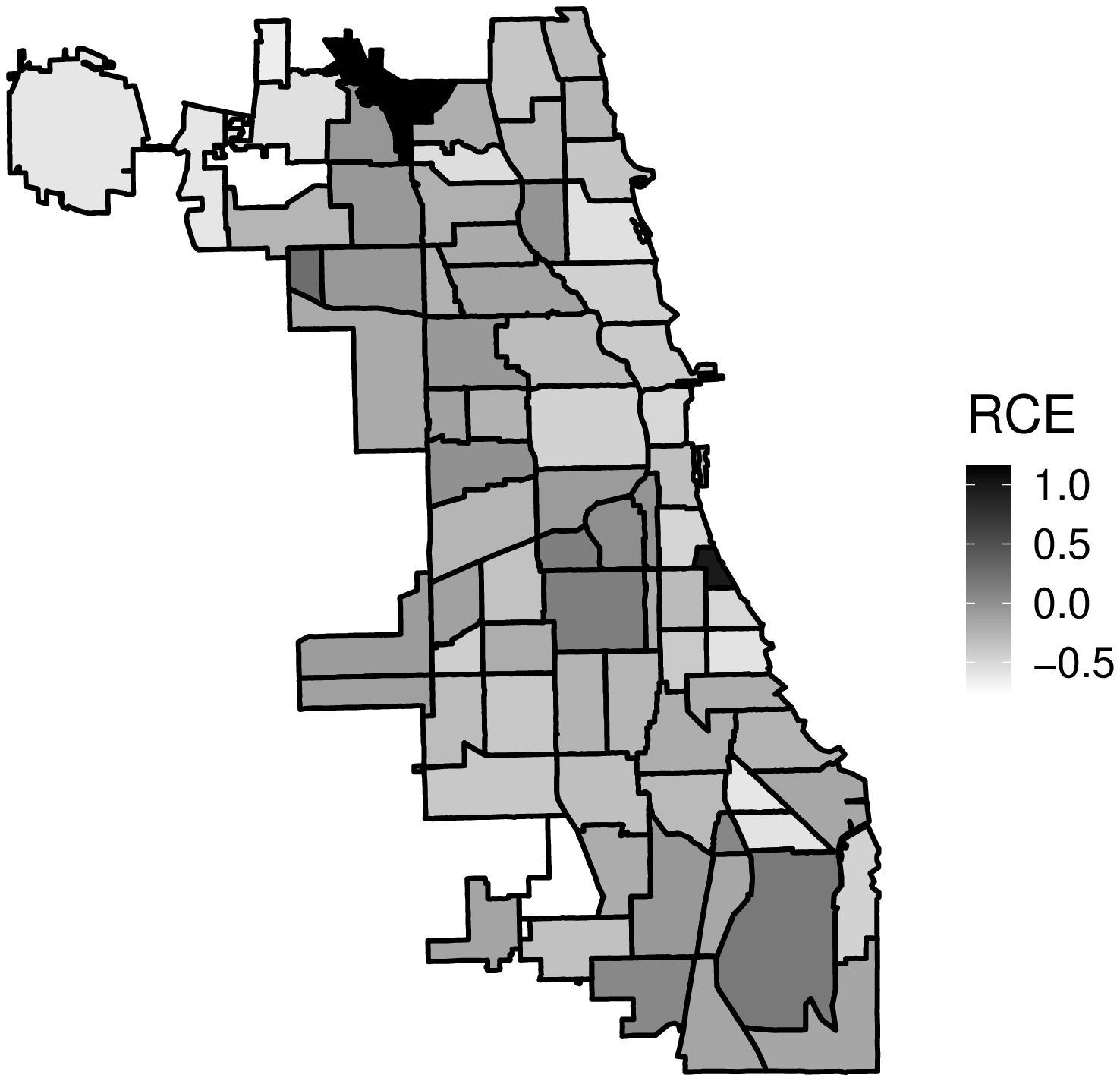}
  \caption{\footnotesize Assaults RCE across Communities}
  \label{fig:sub2}
\end{subfigure}
\caption{}
\label{fig:a1}
\end{figure}

Figure \ref{fig:a1} assesses the changes and RCE distribution for assaults. Compared to burglaries, a higher number of communities experienced significant changes, although within a smaller RCE range (the lowest being an RCE=-0.75). These communities are mostly clustered in the Central, Southwest, and South Side districts. It is worth noting that the literature has shown that the former two districts were generally associated with high to very high levels of crime \citep{Papachristos48YearsCrime2013, SchnellInfluenceCommunityAreas2017}. Nine communities display positive RCE, but only one with a significant effect, with an RCE=1.15.

\begin{figure}[hbt!]
\centering
\begin{subfigure}{0.45\textwidth}
  \centering
  \includegraphics[width=1\linewidth]{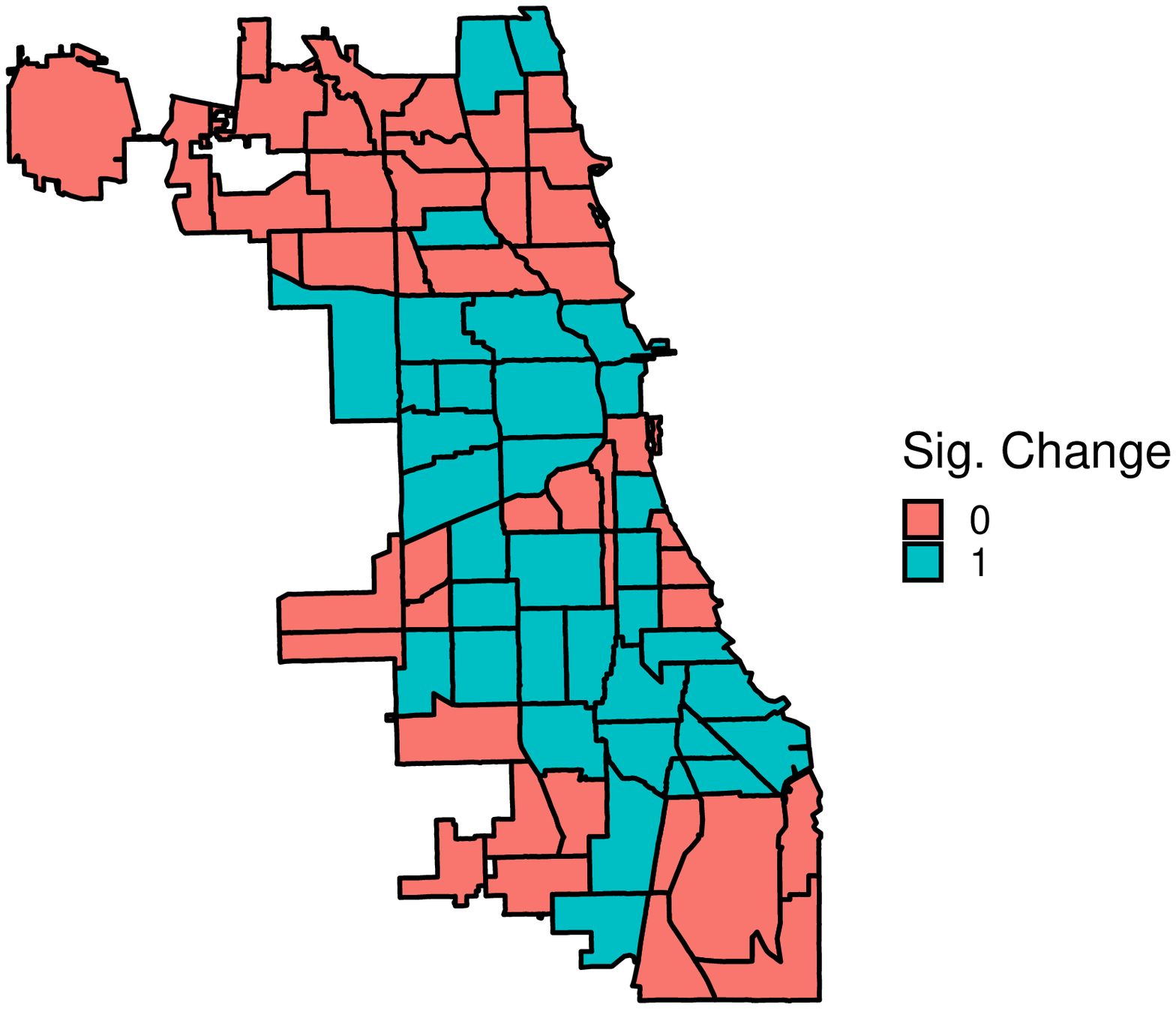}
  \caption{\footnotesize Communities Experiencing Statistically Significant Changes in Narcotics}
  \label{fig:sub1}
\end{subfigure}%
\begin{subfigure}{.45\textwidth}
  \centering
  \includegraphics[width=1\linewidth]{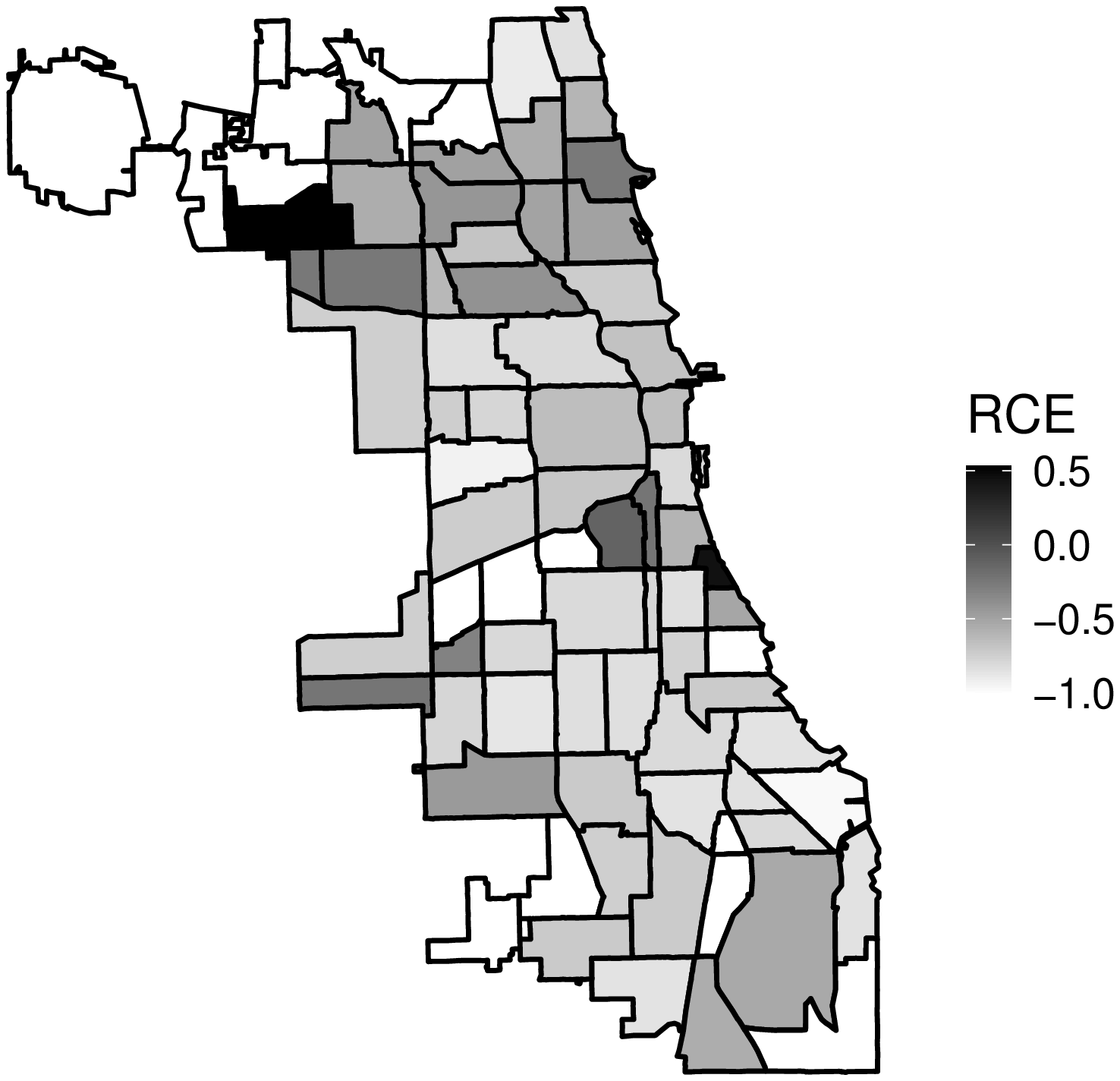}
  \caption{\footnotesize Narcotics RCE across Communities}
  \label{fig:sub2}
\end{subfigure}
\caption{}
\label{fig:n1}
\end{figure}

The community-level situation in terms of narcotics-related crimes is displayed in Figure \ref{fig:n1}. Overall, almost every community experienced a negative RCE: 45.54\% (35 out of 77) of the communities displayed statistically significant reductions in narcotics-related offenses. Most of the communities experiencing significant crime reduction are those that were generally characterized by high levels of crime in the pre-policy period \citep{SchnellInfluenceCommunityAreas2017}, and they are clustered in the West Side, Southwest Side, Far Southeast Side districts. Only 2 out of 77 communities witnessed an increase, but none are significant.

\begin{figure}[hbt!]
\centering
\begin{subfigure}{0.45\textwidth}
  \centering
  \includegraphics[width=1\linewidth]{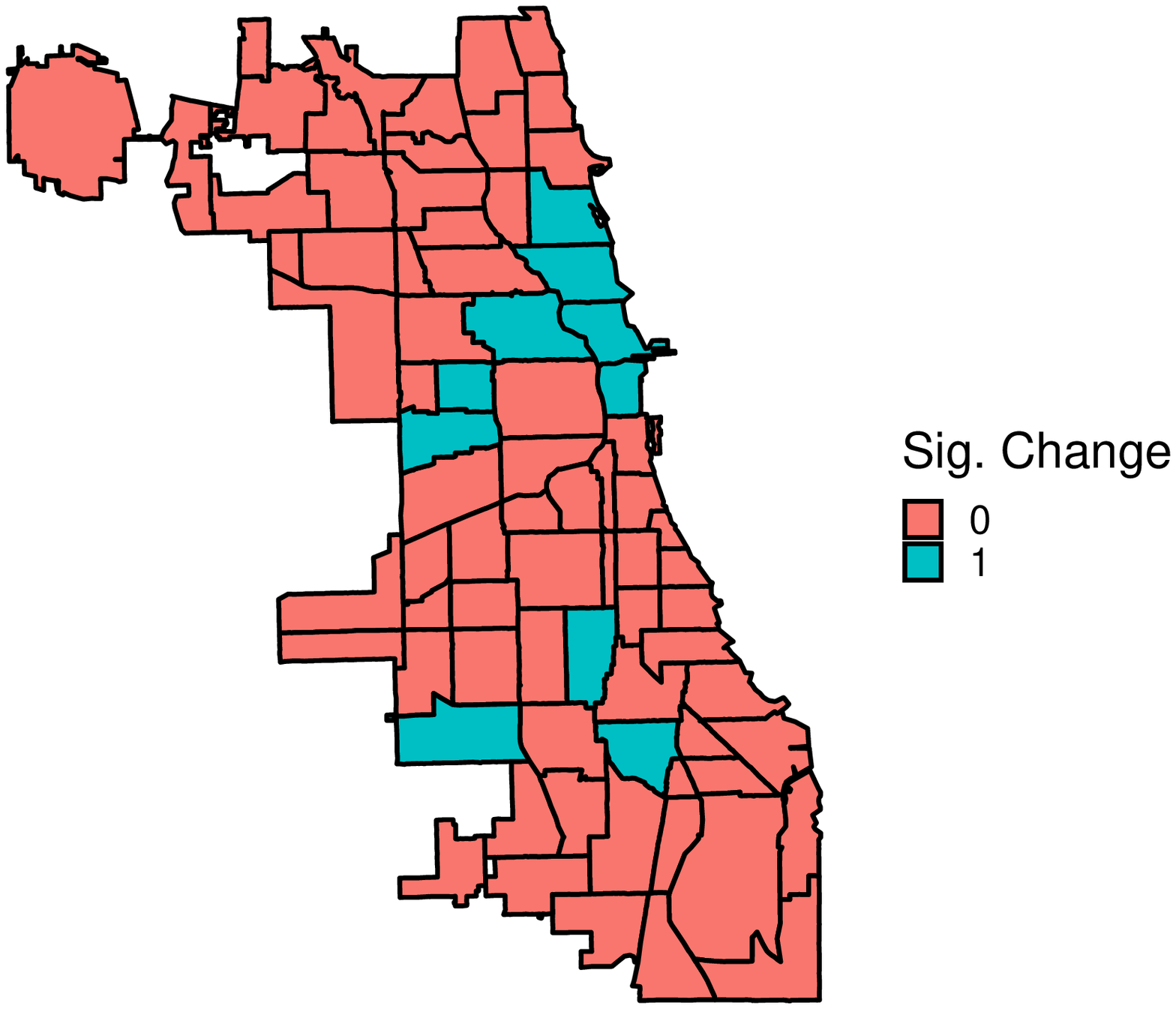}
  \caption{\footnotesize Communities Experiencing Statistically Significant Changes in Robberies}
  \label{fig:sub1}
\end{subfigure}%
\begin{subfigure}{.45\textwidth}
  \centering
  \includegraphics[width=1\linewidth]{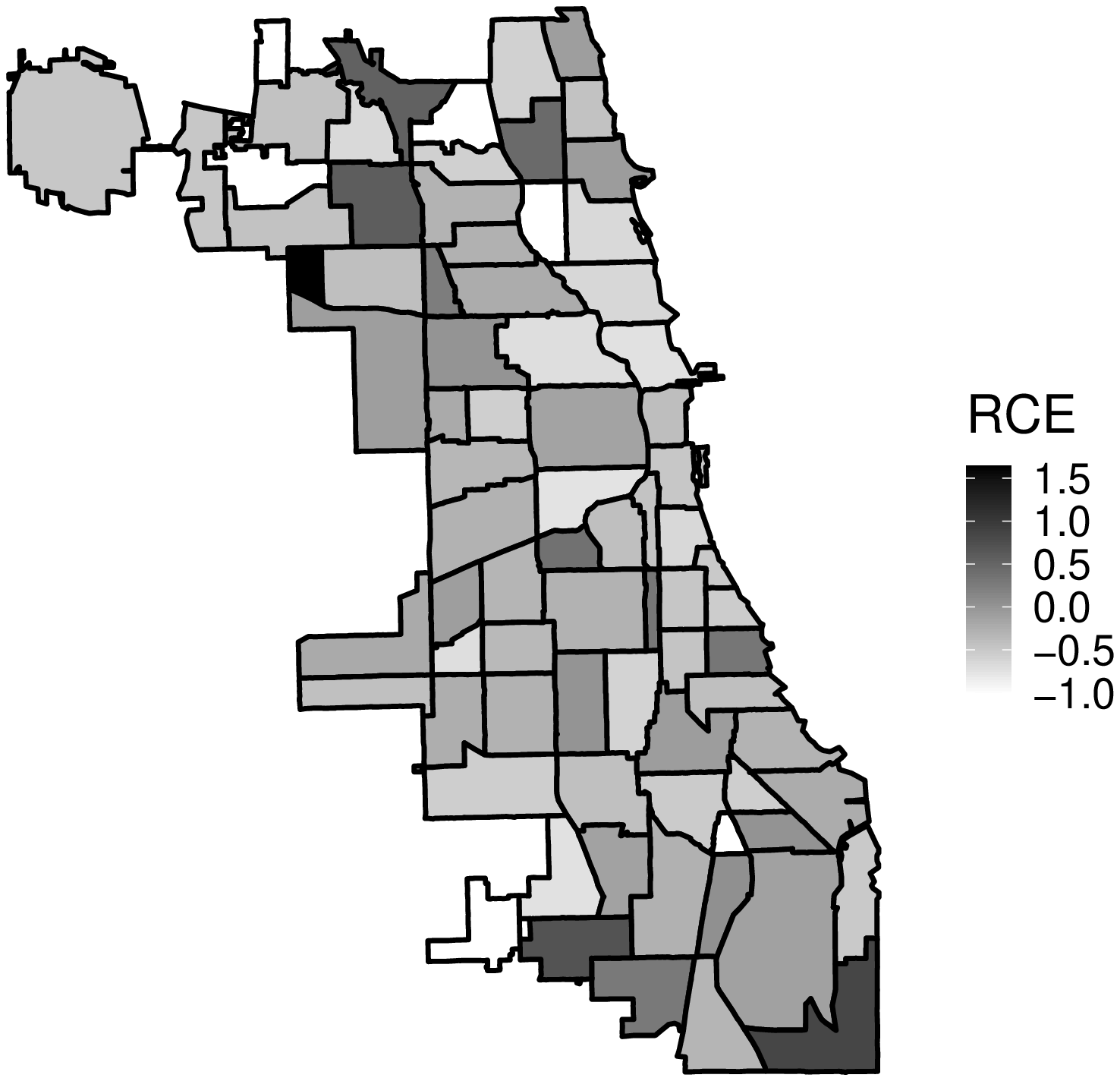}
  \caption{\footnotesize Robberies RCE across Communities}
  \label{fig:sub2}
\end{subfigure}
\caption{}
\label{fig:r1}
\end{figure}

Finally, Figure \ref{fig:r1} highlights the post-intervention effect at the community level in terms of robberies. In a similar fashion to what was found for burglaries, only a minority of communities experience a significant reduction (n=10); these communities are mostly concentrated in the North Side and Central districts. Additionally, 15 communities show an increase in terms of robberies, although none had a positive and statistically significant RCE. 

\begin{figure}[!hbt]
    \centering
    \includegraphics[scale=0.65]{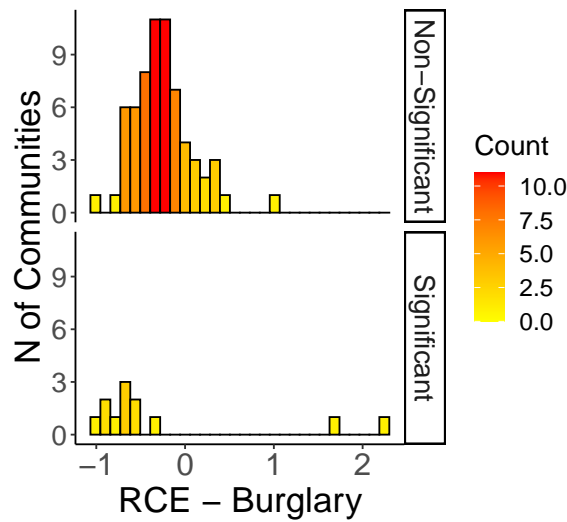}
    \caption{RCE Distribution (Non-Significant vs Significant) – Burglaries}
    \label{fig:burglary_rce}
\end{figure}

\begin{figure}[!hbt]
    \centering
    \includegraphics[scale=0.65]{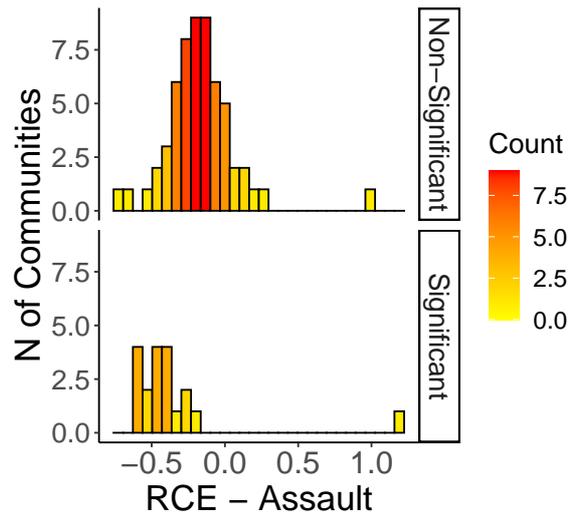}
    \caption{RCE Distribution (Non-Significant vs Significant) – Assaults}
    \label{fig:assault_rce}
\end{figure}

\begin{figure}[!hbt]
    \centering
    \includegraphics[scale=0.65]{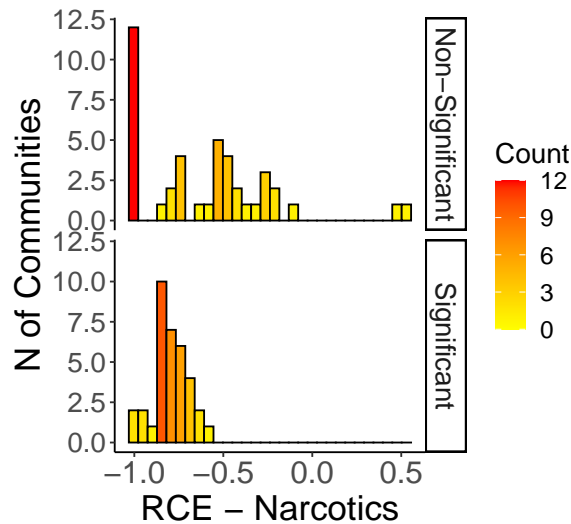}
    \caption{RCE Distribution (Non-Significant vs Significant) – Narcotics}
    \label{fig:narco_rce}
\end{figure}

\begin{figure}[!hbt]
    \centering
    \includegraphics[scale=0.65]{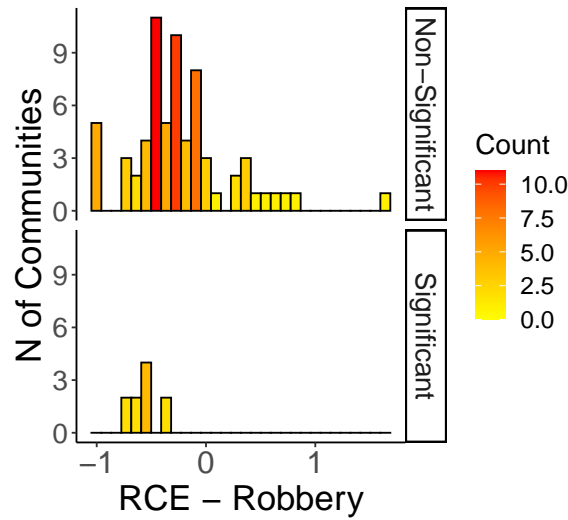}
    \caption{RCE Distribution (Non-Significant vs Significant) – Robberies}
    \label{fig:robbery_rce}
\end{figure}

\color{black} Figures \ref{fig:burglary_rce} through \ref{fig:robbery_rce} show the distribution of RCE for each crime, showing that significant and non-significant values distribute differently. The different distributions allow for drawing some further observations on patterns of RCE. First, significant RCE for burglaries (Figure \ref{fig:burglary_rce}) are clustered in the extreme-left and extreme-right of the histogram, suggesting a correlation between the magnitude and the significance of the effect. Extreme values in assaults are not always significant (Figure \ref{fig:assault_rce}, although also significant changes in assaults tend to cluster around high and low values. The same is true for robberies (Figure \ref{fig:robbery_rce}), and especially for narcotics (Figure \ref{fig:narco_rce}). Robbery and narcotics are also the crime types for which no significant increase is detected; their increases are not statistically significant. \color{black} 

\newpage

\subsection*{Investigating correlates in crime reduction }

Modeling results for each dimension are reported in the following tables. As anticipated in the Methods subsection, for each model, the dependent variable (CR) maps the presence of a significant reduction in the RCE for the given crime type. In light of this, variables displaying positive (and significant) odds ratios are positively associated with the presence of such reduction.

\begin{table}[!hbt]
\footnotesize
\centering
\color{black}
\caption{Firth's Logistic Regression Model for Crime-Related Variables (D.V. is ``Presence of Significant Crime Reduction'' - CR=1)}
\label{crime}

\begin{tabular}{lcccc}
\hline
 & \begin{tabular}[c]{@{}c@{}}(1)\\ Burglary Sig.\\ Reduction\\ OR (SE)\end{tabular} & \begin{tabular}[c]{@{}c@{}}(2)\\ Assault Sig.\\ Reduction\\ OR (SE)\end{tabular} & \begin{tabular}[c]{@{}c@{}}(3)\\ Narcotics Sig. \\ Reduction\\ OR (SE)\end{tabular} & \begin{tabular}[c]{@{}c@{}}(4)\\ Robbery Sig.\\ Reduction\\ OR (SE)\end{tabular} \\ \hline
Neighborhood Safety & \begin{tabular}[c]{@{}c@{}}1.028\\ (0.027)\end{tabular} & \begin{tabular}[c]{@{}c@{}}1.040*\\ (0.024)\end{tabular} & \begin{tabular}[c]{@{}c@{}}0.928***\\ (0.021)\end{tabular} & \begin{tabular}[c]{@{}c@{}}1.028\\ (0.027)\end{tabular} \\
Has Police Station & \begin{tabular}[c]{@{}c@{}}1.130\\ (1.093)\end{tabular} & \begin{tabular}[c]{@{}c@{}}1.247\\ (0.995)\end{tabular} & \begin{tabular}[c]{@{}c@{}}1.387\\ (1.358)\end{tabular} & \begin{tabular}[c]{@{}c@{}}1.316\\ (1.299)\end{tabular} \\
\begin{tabular}[c]{@{}l@{}}Burglary Rate 2019\\ (per 10k in.)\end{tabular} & \begin{tabular}[c]{@{}c@{}}1.010\\ (0.016)\end{tabular} &  &  &  \\
\begin{tabular}[c]{@{}l@{}}Assault Rate 2019\\ (per 10k in.)\end{tabular} &  & \begin{tabular}[c]{@{}c@{}}1.008*\\ (0.005)\end{tabular} &  &  \\
\begin{tabular}[c]{@{}l@{}}Narcotics Rate 2019\\ (per 10k in.)\end{tabular} &  &  & \begin{tabular}[c]{@{}c@{}}1.006\\ (0.005)\end{tabular} &  \\
\begin{tabular}[c]{@{}l@{}}Robbery Rate 2019\\ (per 10k in.)\end{tabular} &  &  &  & \begin{tabular}[c]{@{}c@{}}1.020*\\ (0.011)\end{tabular} \\ 
Intercept & \begin{tabular}[c]{@{}c@{}}0.014*\\ (0.036) \end{tabular}  & \begin{tabular}[c]{@{}c@{}}0.007*\\ (0.016) \end{tabular} & \begin{tabular}[c]{@{}c@{}}114.530***\\ (234.024)\end{tabular} & \begin{tabular}[c]{@{}c@{}}0.008*\\ (0.020)\end{tabular}\\\hline
N & 77 & 77 & 77 & 77 \\
Chi2 & 1.073 & 3.355 & 16.826 & 3.257 \\
p & 0.783 & 0.340 & 0.001 & 0.354 \\ \hline
\end{tabular}

{****Significant at 99.9\%, ***Significant at 99\%, **Significant at 95\%, *Significant at 90\%}
\end{table}
\color{black}
For each crime, Table \ref{crime} shows the effects of the crime-related correlates described above. \color{black}
The presence of a significant reduction in burglaries is not predicted by any of the included correlates; not even the rate of burglaries that occurred in 2019 is associated with a decrease in the post-intervention period. Conversely, effects are found for other crimes. The rate of assaults computed for 2019 is positively associated with a statistically significant reduction of assaults in the post-intervention period (OR=1.008, p$<$0.1). The same pattern is found for robberies: the higher the rate of robberies in 2019, the higher the likelihood of a significant reduction after the introduction of the containment policies (OR=1.020, p$<$0.1).

Although the two effect sizes seem negligible in absolute terms, they are instead relevant considering the variation of assaults and burglaries across Chicago communities. The rate of assaults in 2019 has a mean value of 87.84, with a range going from a minimum of 14.01 to a maximum of 297.37 (St. Dev.= 69.19). The burglary rate also indicates high variability: the average is 38.18, going from a minimum of 5.96 to a maximum of 110.45 (St. Dev.= 23.84).

Reductions in narcotics-related offenses are successfully predicted by perceived neighborhood safety only (OR=0.928, p$<$0.01). Specifically, the relationship indicates that the higher the perceived safety in a neighborhood the lower the odds of reduction in narcotics-related offenses. This finding might be related to policing dynamics: police may decide to target communities that are generally less safe, hence reducing the possibility to positively influence narcotics activity in safer areas. Conversely, safer neighborhoods have higher odds to experience reductions in assaults (OR=1.040, p$<$0.1). This finding may be explained by the fact that social distancing and shelter-in-place policies further reduce the opportunity for assaults in already safe communities by sensibly reducing the number of potential victims in the streets.  Lastly, the presence of a police station within a community does not have any significant impact on crime reductions.\footnote{Since data on police stations were gathered with their associated ZIP codes, and since a ZIP code may belong to more than one community in the city of Chicago, a community was labeled as “has police” if there was at least one police station present in a census associated to the community. If, for instance, community A and B share ZIP code y, and ZIP code y has a police station, both A and B are labeled as “has police” communities.} While the magnitude of the OR is mostly positive (and high), except for burglaries, the statistical relationship is not strong enough to make any inference about the deterrence role of police stations in the post-intervention timeframe.  

\color{black}
\begin{table}[!hbt]
\footnotesize
\centering
\color{black}
\caption{Firth's Logistic Regression Model for Socio-Economic Variables  (D.V. is ``Presence of Significant Crime Reduction'' - CR=1)}

\begin{tabular}{lcccc}
\hline
 & \begin{tabular}[c]{@{}c@{}}(1)\\ Burglary Sig.\\ Reduction\\ OR (SE)\end{tabular} & \begin{tabular}[c]{@{}c@{}}(2)\\ Assault Sig.\\ Reduction\\ OR (SE)\end{tabular} & \begin{tabular}[c]{@{}c@{}}(3)\\ Narcotics Sig. \\ Reduction\\ OR (SE)\end{tabular} & \begin{tabular}[c]{@{}c@{}}(4)\\ Robbery Sig.\\ Reduction\\ OR (SE)\end{tabular} \\ \hline
Crowded Housing (\%) & \begin{tabular}[c]{@{}c@{}}1.204\\ (0.148)\end{tabular} & \begin{tabular}[c]{@{}c@{}}0.923\\ (0.146)\end{tabular} & \begin{tabular}[c]{@{}c@{}}1.140\\ (0.135)\end{tabular} & \begin{tabular}[c]{@{}c@{}}0.755\\ (0.175)\end{tabular} \\
Vacant Housing (\%) & \begin{tabular}[c]{@{}c@{}}0.967\\ (0.099)\end{tabular} & \begin{tabular}[c]{@{}c@{}}0.905\\ (0.102)\end{tabular} & \begin{tabular}[c]{@{}c@{}}1.318***\\ (0.127)\end{tabular} & \begin{tabular}[c]{@{}c@{}}0.954\\ (0.104)\end{tabular} \\
Income Diversity & \begin{tabular}[c]{@{}c@{}}0.810*\\ (0.094)\end{tabular} & \begin{tabular}[c]{@{}c@{}}1.446**\\ (0.246)\end{tabular} & \begin{tabular}[c]{@{}c@{}}1.270***\\ (0.133)\end{tabular} & \begin{tabular}[c]{@{}c@{}}0.744**\\ (0.081)\end{tabular} \\
Poverty Rate & \begin{tabular}[c]{@{}c@{}}0.833**\\ (0.055)\end{tabular} & \begin{tabular}[c]{@{}c@{}}1.066\\ (0.060)\end{tabular} & \begin{tabular}[c]{@{}c@{}}1.083\\ (0.055)\end{tabular} & \begin{tabular}[c]{@{}c@{}}0.976\\ (0.061)\end{tabular} \\
Total Population (\textbackslash{}1000) & \begin{tabular}[c]{@{}c@{}}1.039**\\ (0.016)\end{tabular} & \begin{tabular}[c]{@{}c@{}}1.039***\\ (0.015)\end{tabular} & \begin{tabular}[c]{@{}c@{}}1.032**\\ (0.014)\end{tabular} & \begin{tabular}[c]{@{}c@{}}1.056***\\ (0.019)\end{tabular} \\
\begin{tabular}[c]{@{}l@{}}Burglary Rate 2019\\ (per 10k in.)\end{tabular} & \begin{tabular}[c]{@{}c@{}}1.030\\ (0.024)\end{tabular} &  &  &  \\
\begin{tabular}[c]{@{}l@{}}Assault Rate 2019\\ (per 10k in.)\end{tabular} &  & \begin{tabular}[c]{@{}c@{}}1.029**\\ (0.013)\end{tabular} &  &  \\
\begin{tabular}[c]{@{}l@{}}Narcotics Rate 2019\\ (per 10k in.)\end{tabular} &  &  & \begin{tabular}[c]{@{}c@{}}1.003\\ (0.006)\end{tabular} &  \\
\begin{tabular}[c]{@{}l@{}}Robbery Rate 2019\\ (per 10k in.)\end{tabular} &  &  &  & \begin{tabular}[c]{@{}c@{}}1.013\\ (0.012)\end{tabular} \\
Intercept & \begin{tabular}[c]{@{}c@{}}3.10e+4\\ (2.33e+5)\end{tabular} & \begin{tabular}[c]{@{}c@{}}0.000**\\ (0.000)\end{tabular} & \begin{tabular}[c]{@{}c@{}}0.000*\\ (0.000)\end{tabular} & \begin{tabular}[c]{@{}c@{}}4.35e+5\\ (3.02e+6)\end{tabular} \\ \hline
N & 77 & 77 & 77 & 77 \\
Chi2 & 8.384 & 10.066 & 17.579 & 13.459 \\
p & 0.211 & 0.122 & 0.007 & 0.036 \\ \hline
\end{tabular}
\label{socio}

{****Significant at 99.9\%, ***Significant at 99\%, **Significant at 95\%, *Significant at 90\%}
\end{table}
\color{black}
Table \ref{socio} focuses on the socio-economic variables. Two relations are found concerning the significant reduction of burglaries. First, more populated communities are more prone to experience significant reductions. The positive relation between significant crime reduction and population is found for all the considered crimes (burglary OR=1.039, assault OR=1.039, narcotics OR=1.032, robbery OR=1.056). In fact, the computation of the average marginal effect shows that burglaries are expected to reduce by 3.8\% (OR=1.039) when the population of the average community increases by 1 unit (an increase of 1,000 inhabitants), and other factors are held constant at their mean. Still, the actual effects can be relevant given that communities in Chicago are heterogeneous in terms of population (the minimum is 2,254 and the maximum is 100,470 with a standard deviation of 22,916.61). We interpret the overall effect of population in the frame of Routine Activity Theory: more inhabitants forced at home enhance guardianship, reducing crime opportunities \citep{CohenSocialChangeCrime1979}.

Contrarily, higher levels of poverty are negatively associated with the dependent variable, in line with previous research on the effect of low economic conditions and crime \citep{PareIncomeinequalitypoverty2014}. If the rate of inhabitants living in poverty in an average community (mean 21.97) increases by a 1 percentage point, the predicted probability of a significant reduction in burglaries decreases by 12.4\% (OR=0.833, p$<$0.05), as per computation of the Average Marginal Effect.

As we are focusing on crime reductions, our initial hypothesis was the odds of detecting crime reduction to decrease with each incremental rise in house crowding. Yet, contrary to the theoretical premises of General Strain Theory, crowded housing is not associated with any of the variables under consideration, not even with assaults.

The odds of significant narcotics reductions increase by a factor of 1.318 times for each additional increment of vacant housing units, contrary to what we might have expected based on the literature \citep{SpelmanAbandonedbuildingsMagnets1993}. This suggests that the stay at home order and social distancing policy recommendations may have discouraged people from moving around the city to commit crimes in communities with a high prevalence of unguarded buildings. 

Income diversity shows divergent effects. On one hand, a positive relation between income diversity and significant crime reductions is detected for narcotics (OR=1.270, p$<$0.05) and assaults (OR=1.446, p$<$0.05). On the other hand, a unit increase in income diversity reduces the odds of witnessing significant reductions in burglaries (OR=0.810, p$<$0.1) and robberies (OR=0.744, p$<$0.05). This might indicate that economically heterogeneous structures of communities create the conditions for palatable robbery and burglary targets also in a condition of highly disrupted human mobility. An underlying factor associated with this pattern can be relative deprivation, which is connected in turn to resentment and frustration \citep{AgnewGeneralStrainTheory1999, WilkinsonWhyviolencemore2004}.

\begin{table}[!hbt]
\footnotesize
\centering
\color{black}
\caption{Firth's Logistic Regression Model for Health/Demographic Variables  (D.V. is ``Presence of Significant Crime Reduction'' - CR=1)}

\begin{tabular}{lcccc}
\hline
 & \begin{tabular}[c]{@{}c@{}}(1)\\ Burglary Sig.\\ Reduction\\ OR (SE)\end{tabular} & \begin{tabular}[c]{@{}c@{}}(2)\\ Assault Sig.\\ Reduction\\ OR (SE)\end{tabular} & \begin{tabular}[c]{@{}c@{}}(3)\\ Narcotics Sig. \\ Reduction\\ OR (SE)\end{tabular} & \begin{tabular}[c]{@{}c@{}}(4)\\ Robbery Sig.\\ Reduction\\ OR (SE)\end{tabular} \\ \hline
\% Population Aged \textgreater{}65 & \begin{tabular}[c]{@{}c@{}}0.781**\\ (0.087)\end{tabular} & \begin{tabular}[c]{@{}c@{}}1.008\\ (0.074)\end{tabular} & \begin{tabular}[c]{@{}c@{}}0.863**\\ (0.067)\end{tabular} & \begin{tabular}[c]{@{}c@{}}0.782**\\ (0.091)\end{tabular} \\
\% Population Aged \textless{}18 & \begin{tabular}[c]{@{}c@{}}0.932\\ (0.063)\end{tabular} & \begin{tabular}[c]{@{}c@{}}0.862**\\ (0.052)\end{tabular} & \begin{tabular}[c]{@{}c@{}}0.973\\ (0.049)\end{tabular} & \begin{tabular}[c]{@{}c@{}}0.957\\ (0.061)\end{tabular} \\
Overall Health Status & \begin{tabular}[c]{@{}c@{}}1.015\\ (0.052)\end{tabular} & \begin{tabular}[c]{@{}c@{}}1.026\\ (0.045)\end{tabular} & \begin{tabular}[c]{@{}c@{}}0.919**\\ (0.035)\end{tabular} & \begin{tabular}[c]{@{}c@{}}1.178**\\ (0.092)\end{tabular} \\
\begin{tabular}[c]{@{}l@{}}ln(Covid Cases Rate)\\ (per 10k in)\end{tabular} & \begin{tabular}[c]{@{}c@{}}1.130\\ (0.631)\end{tabular} & \begin{tabular}[c]{@{}c@{}}0.568\\ (0.238)\end{tabular} & \begin{tabular}[c]{@{}c@{}}0.477*\\ (0.206)\end{tabular} & \begin{tabular}[c]{@{}c@{}}0.633\\ (0.335)\end{tabular} \\
\begin{tabular}[c]{@{}l@{}}Burglary Rate 2019\\ (per 10k in.)\end{tabular} & \begin{tabular}[c]{@{}c@{}}1.019\\ (0.018)\end{tabular} &  &  &  \\
\begin{tabular}[c]{@{}l@{}}Assault Rate 2019\\ (per 10k in.)\end{tabular} &  & \begin{tabular}[c]{@{}c@{}}1.010*\\ (0.005)\end{tabular} &  &  \\
\begin{tabular}[c]{@{}l@{}}Narcotics Rate 2019\\ (per 10k in.)\end{tabular} &  &  & \begin{tabular}[c]{@{}c@{}}1.030***\\ (0.011)\end{tabular} &  \\
\begin{tabular}[c]{@{}l@{}}Robbery Rate 2019\\ (per 10k in.)\end{tabular} &  &  &  & \begin{tabular}[c]{@{}c@{}}1.053****\\ (0.016)\end{tabular} \\
Intercept & \begin{tabular}[c]{@{}c@{}}1.111\\ (6.529)\end{tabular} & \begin{tabular}[c]{@{}c@{}}12.307\\ (67.700)\end{tabular} & \begin{tabular}[c]{@{}c@{}}3.04e+5\\ (1.54e+6)\end{tabular} & \begin{tabular}[c]{@{}c@{}}0.000\\ (0.000)\end{tabular} \\ \hline
N & 77 & 77 & 77 & 77 \\
Chi2 & 5.657 & 11.877 & 12.837 & 13.096 \\
p & 0.341 & 0.037 & 0.025 & 0.022 \\ \hline
\end{tabular}
\label{hea}

{****Significant at 99.9\%, ***Significant at 99\%, **Significant at 95\%, *Significant at 90\%}
\end{table}
\color{black}
Concerning demographic and health-associated variables, heterogeneous associations emerge. First of all, the presence of elderly inhabitants is found to be significantly and negatively associated only in terms of a significant reduction in burglary (Table \ref{hea}). 

Specifically, a unit increase in the rate of residents aged 65 or more leads to a 21.9\% reduction in the likelihood of having a significant reduction in burglaries. This is in line with the hypothesis that burglars may target areas with a high share of elderly individuals assuming lower levels of capable guardianship or physical resistance. In line with this finding, also a reduction in robberies is negatively associated with the presence of seniors in a given community (OR=0.782, p$<$0.05). This further suggests that offenders may have decided to specifically target strata of the population with weaker resistance capabilities.

Assaults display instead a statistically significant negative relationship with the prevalence of juveniles in the community. The odds ratio of 0.863 means that, following the computation of the average marginal effect, an increase of 1 percentage point in the rate of inhabitants aged less than 18 years in the average community (mean 22.5) leads to a 14.9\% reduction in the probability of a significant reduction of assaults. This finding can be linked to both the literature about the effect of age structure on aggregate levels of crime and research in the context of life-course criminology \citep{CohenAgeStructureCrime1987, ShavitAgeCrimeEarly1988, SweetenAgeExplanationCrime2013}, but also to the fact that youth may be less prone to comply with the containment interventions \citep{MooreExperienceSocialDistancing2020, GowenYoungpeopleurged}.\color{black}

The prevalence of COVID-19 cases in a community does not affect the presence of a significant crime reduction on three crime categories out of four, the ones in which there is a victim. The only exception is narcotics; a unitary increase in the log-transformed rate of cases is negatively associated with a decline in drug-related crimes (OR=0.477, p$<$0.1). This relation may be driven by lower compliance with formal norms that simultaneously applies to both drug consumption and violation of social distancing measures \citep{JackaDrugusepandemic2020, DumasWhatDoesAdolescent2020} \color{black} which, in turn, manifests in a higher prevalence of COVID-19 cases and a lower reduction in narcotics.

The overall health status variable, which measures the “share of the population of adults aged 18 years and older who reported that their overall health is good, very good or excellent” is significantly and negatively associated only with a reduction in narcotics. Specifically, a unit increase in this share leads to an 8.1\% reduction in the likelihood of witnessing a statistically significant reduction of drug-related offenses. Contrarily, the relationship is the opposite in terms of robberies: the higher the perceived health status, the higher the likelihood of a significant reduction in the post-intervention period (OR=1.178, p$<$0.05).

\begin{table}[!hbt]
\footnotesize
\centering
\color{black}
\caption{Firth's Logistic Regression Model for Other Crimes Evolution (D.V. is ``Presence of Significant Crime Reduction'' - CR=1)}
\begin{tabular}{lcccc}
\hline
 & \begin{tabular}[c]{@{}c@{}}(1)\\ Burglary Sig.\\ Reduction\\ OR (SE)\end{tabular} & \begin{tabular}[c]{@{}c@{}}(2)\\ Assault Sig.\\ Reduction\\ OR (SE)\end{tabular} & \begin{tabular}[c]{@{}c@{}}(3)\\ Narcotics Sig. \\ Reduction\\ OR (SE)\end{tabular} & \begin{tabular}[c]{@{}c@{}}(4)\\ Robbery Sig.\\ Reduction\\ OR (SE)\end{tabular} \\ \hline
Sig. Reduction Assaults & \begin{tabular}[c]{@{}c@{}}0.978\\ (0.768)\end{tabular} &  & \begin{tabular}[c]{@{}c@{}}3.024*\\ (1.851)\end{tabular} & \begin{tabular}[c]{@{}c@{}}2.123\\ (1.593)\end{tabular} \\
Sig. Reduction Narcotics & \begin{tabular}[c]{@{}c@{}}1.958\\ (1.958)\end{tabular} & \begin{tabular}[c]{@{}c@{}}2.984*\\ (1.920)\end{tabular} &  & \begin{tabular}[c]{@{}c@{}}1.512\\ (1.122)\end{tabular} \\
Sig. Reduction Robberies & \begin{tabular}[c]{@{}c@{}}5.801**\\ (4.417)\end{tabular} & \begin{tabular}[c]{@{}c@{}}2.046\\ (1.528)\end{tabular} & \begin{tabular}[c]{@{}c@{}}1.413\\ (1.323)\end{tabular} &  \\
Sig. Reduction Burglaries &  & \begin{tabular}[c]{@{}c@{}}1.025\\ (0.799)\end{tabular} & \begin{tabular}[c]{@{}c@{}}2.494\\ (2.023)\end{tabular} & \begin{tabular}[c]{@{}c@{}}6.373**\\ (4.971)\end{tabular} \\
\begin{tabular}[c]{@{}l@{}}Burglary Rate 2019\\ (per 10k in.)\end{tabular} & \begin{tabular}[c]{@{}c@{}}0.987\\ (0.018)\end{tabular} &  &  &  \\
\begin{tabular}[c]{@{}l@{}}Assault Rate 2019\\ (per 10k in.)\end{tabular} &  & \begin{tabular}[c]{@{}c@{}}0.988\\ (0.005)\end{tabular} &  &  \\
\begin{tabular}[c]{@{}l@{}}Narcotics Rate 2019\\ (per 10k in.)\end{tabular} &  &  & \begin{tabular}[c]{@{}c@{}}1.024***\\ (0.009)\end{tabular} &  \\
\begin{tabular}[c]{@{}l@{}}Robbery Rate 2019\\ (per 10k in.)\end{tabular} &  &  &  & \begin{tabular}[c]{@{}c@{}}1.015**\\ (0.008)\end{tabular} \\
Intercept & \begin{tabular}[c]{@{}c@{}}0.139***\\ (0.098)\end{tabular} & \begin{tabular}[c]{@{}c@{}}0.193****\\ (0.095)\end{tabular} & \begin{tabular}[c]{@{}c@{}}0.204****\\ (0.092)\end{tabular} & \begin{tabular}[c]{@{}c@{}}0.039****\\ (0.029)\end{tabular} \\ \hline
N & 77 & 77 & 77 & 77 \\
Chi2 & 6.721 & 4.916 & 11.385 & 9.889 \\
p & 0.151 & 0.296 & 0.023 & 0.042 \\ \hline
\end{tabular}
\label{joi}

{****Significant at 99.9\%, ***Significant at 99\%, **Significant at 95\%, *Significant at 90\%}
\end{table}
\color{black}
The joint reduction models show that there are two pairwise relationships, but we do not observe offense displacement; the correlation between crime reductions are positive (Table \ref{joi}). A significant reduction in robberies is associated with higher odds of a reduction in burglaries in the same community (OR=5.801, p$<$0.05). In parallel, the odds of a significant reduction in robberies, in a community where burglaries reduced as well, are 6.373 (p$<$0.05). This suggests that the two crimes are jointly influenced by common underlying opportunity structures that exist in the communities. Similarly, a significant reduction in assaults is associated with a significant reduction in narcotics (OR=2.984, p$<$0.10), and a significant reduction in narcotics is associated with a joint significant reduction in assaults (OR=3.024, p$<$0.10). Also, in this case, this means that regardless of the single determinants/risk factors previously assessed--which are often different for the different crime types--there are other dynamics, which associate these two pairs of crime types.
\color{black} 

\section*{Discussion}
In line with the literature on crime concentration, we found that the issued COVID-19 containment policies have impacted crime in different ways across communities. In terms of burglaries, only 10 communities (12.98\% of the total) experienced a statistically significant reduction during our post-intervention follow-up period. Concerning assaults, 18 communities (23.37\%) experienced a significant reduction. Narcotics-related offenses are those that significantly declined in more areas compared to the other crimes, with 35 communities associated with a statistical reduction (45.54\%). Finally, robberies have been significantly reduced in only 10 communities (12.98\%).

We also detected cases in which crimes have significantly increased. Significantly more burglaries than expected occurred in 2 communities (with increases of 219\% and 168\%, respectively) and 1 community experienced a statistically significant increase of assaults (+115\%). \color{black}As the overall number of assaults has contracted after the introduction of containment policies, the increase registered at Forest Glen may relate to a spatial displacement dynamic. The forms of displacement related to the increase in burglaries might, instead, be two; in addition to spatial displacement also offense displacement--i.e., from residential burglaries to non-residential burglaries--might have happened. Displacement relates to dissimilar changes in the crime environment. While differences in criminal opportunities related to residential and non-residential burglaries maybe reconducted to a higher--or lower--guardianship during the stay-at-home period, other modifications in the criminal environment would need to be addressed with ad-hoc empirical strategies. More in general, evidence pointing in the direction of increases in crime--although circumscribed to few communities and few crimes--suggests further researches aiming at explaining them should be conducted.\color{black}

For what concerns the inferential models investigating correlates of statistically significant reductions, mixed findings emerged. \color{black} The 2019 level of assaults was positively associated with the statistical reduction of the same reference crime category. The same applies to robberies. This suggests that communities that had previously high levels of such offenses benefited more from the COVID-19 containment policies, compared to other communities. Concerning perceived neighborhood safety, two opposite relationships were found. Higher perceived safety in a community predicts significant reductions in assaults, while a contrary link was found for narcotics, possibly due to policing dynamics. Finally, the presence of a police station located in the ZIP codes that are part of a community did not lead to any significant association with crime reduction. 

From the socio-economic point of view, the rate of people living in crowded houses was not found to be negatively associated with statistical reductions in assaults, contrary to the theoretical premises of GST \citep{FreedmanCrowdingbehavior1975, BoothCrowdingUrbancrime1976, AgnewFoundationGeneralStrain1992}. Also surprisingly, vacant housing rate was positively associated with significant reductions in narcotics. Income diversity was positively associated with reductions in narcotics-related offenses and assaults but negatively associated with a decline in burglaries and assaults. This latter result may be linked to the role of relative deprivation in shaping criminal activity within economically heterogeneous communities \citep{BlauCostInequalityMetropolitan1982}. In parallel, poverty is negatively associated with reductions in burglaries suggesting that absolute deprivation, in general, has a negative impact--especially in such an exceptional condition--on appropriative crimes. The number of inhabitants is the only variable positively associated with all the considered crime types, indicating that the higher the number of people at home, the higher the levels of capable guardianship, the lower the opportunities for crime. 

Indications also emerged from the analysis of health-related and demographic variables. First, the higher the share of people aged more than 65 years old, the lower the odds of observing a statistically significant reduction of burglaries. This can be related to the fact that burglars prefer to target people that pose weaker resistance and guardianship to their residencies \color{black}and that retired people tend to spend more time at home than younger ones; therefore, containment policies might introduce smaller changes in the communities where older people are more numerous. Furthermore, the higher the number of people aged less than 18, the higher the levels of assaults, in line with perspectives on life-course criminology and with the hypothesis that youth are less prone to comply with rules and restrictions.

The prevalence of COVID-19 infections in a community did not lead to any effect on crime. Self-perceived health status was associated with reductions in robberies, while the relationship was inverse for narcotics. A possible explanation of this dynamic lies in the fact that a higher perception of health status might be related to higher compliance with mobility restriction orders \citep{BarariEvaluatingCOVID19Public2020} \color{black}. In turn, the higher level of compliance induces a reduction in robberies. Conversely, drug consumption may associate with lower self-perceived health, possibly via worse health status \citep{NealeMeasuringhealthScottish2004, MillsonSelfperceivedHealthCanadian2004}\color{black}.

Finally, we have found evidence of two pairwise relations between joint reduction in two crime types of appropriative nature: observing a reduction in burglaries was strongly associated with the odds of observing also a reduction in robberies, and vice-versa. Additionally, observing a reduction in narcotics in a community is associated with very high odds of observing a reduction also in assaults and vice versa. 
\newpage
\section*{\color{black} Conclusions}
\color{black}This work analyzed the community-level trends in burglaries, assaults, narcotics-related offenses and robberies to understand whether and how COVID-19 containment policies have impacted differently across different areas of the same city, and what are the main contextual correlates associated to crime reductions. We framed our work in the context of Routine Activity Theory \citep{CohenSocialChangeCrime1979}, Crime Pattern Theory \citep{BrantinghamPatternsCrime1984} and General Strain Theory \citep{AgnewFoundationGeneralStrain1992, AgnewGeneralStrainTheory1999}, but the community-level focus of the analyses are motivated by the literature on crime concentration which indicates that crime does not occur randomly in time and space \citep{ShawJuveniledelinquencyurban1942, FreemanSpatialConcentrationCrime1996b, Johnsonbriefhistoryanalysis2010, WeisburdLawCrimeConcentration2015a}. It rather clusters in certain areas of a city. Concerning our crimes of interest, Routine Activity Theory and Crime Pattern Theory both point in the direction of diffused decrease in crime events due to the massive reduction in opportunities (a joint effect of increased capable guardianship and very low human mobility). On the other hand, General Strain Theory suggests that the presence of negative stimuli--such as the stress associated to isolation and financial and economic uncertainty--and the impossibility to reach positively valued goals due to the stay-at-home policies may lead to crime by increasing negative emotions responses, thus predicting potential spikes in certain types of offenses. As these propositions govern the overall dynamics concerning crime, we posited that such theories can explain the extent to which criminal activity varies across Chicago communities due to their different criminal, social, economic, health, and demographic contexts. Indeed, certain areas of the city experienced crime reduction, while others did not. These reductions are crime-sensitive and depend on the different contextual characteristics of the communities.

Although providing new insights on the relation between the COVID-19 pandemic and crime, our work comes with some limitations. First, there is a well-known gap between actual and reported crimes. It is thus important to be careful when interpreting inferential results that are solely based on reported events. In addition, crimes can also be reported later compared to the day when they were actually committed. To partially mitigate possible biases coming from short-term late reporting we collected the data on June 06th 2020–three weeks after May 17th 2020 which is the last day of our time frame. Nonetheless, this does not help to solve problems related to long-term late reporting. Further, the literature on crime and place has consistently shown that crime is patterned at micro-levels. This means that going more in detail at the geographical level may reveal further mechanics and patterns that our analyses fail to detect and consider. Community-wide analyses already disentangle diverging dynamics, but Chicago’s communities are highly complex, multifaceted, and diverse, and it is thus likely that more micro approaches would be useful in highlighting further dynamics. Thirdly, we only considered a small set of crimes and some aggregate crime categories (i.e, burglary does not distinguish between residential and non-residential crimes). Given the heterogeneity of the results we provided, we expect many more layers of complexity intervening in all the other offenses that we have not considered. Future work should address this gap.

Despite these limitations, the many outcomes of our work first demonstrate that crime trends do not behave in the same way across different areas of the same city and across different types of crimes in the context of the pandemic. Additionally, we show that there are different factors connected to community-level reductions in crimes. This finding prompts new research questions and provides relevant insights also from a policy standpoint. The inhomogeneous distribution and magnitude of effects and the statistical associations found with variables mapping the context of each Chicago community can help inform public policies and police practices aimed at preventing and tackling crime in this exceptional situation. This will be extremely important moving forward especially as restrictions are eased and then again if—and when—another wave of the virus breaks out and policies are re-introduced.

\subsection*{Code and Data Availability}
Code and data are available at \texttt{https://github.com/gcampede}. For further information please contact the corresponding author.
\subsection*{Funding}
The authors have not received any funding for the present work.
\subsection*{Competing Interests}
The authors declare no competing interests.

\newpage
\bibliographystyle{apalike} % Style BST file (bmc-mathphys, vancouver, spbasic).
\bibliography{chicago_covid.bib}

\begin{thebibliography}{}

\bibitem[Abrams, 2020]{AbramsCOVIDCrimeEarly2020}
Abrams, D. (2020).
\newblock {COVID} and {Crime}: {An} {Early} {Empirical} {Look}.
\newblock {SSRN} {Scholarly} {Paper} ID 3674032, Social Science Research
  Network, Rochester, NY.

\bibitem[Agnew, 1992]{AgnewFoundationGeneralStrain1992}
Agnew, R. (1992).
\newblock Foundation for a {General} {Strain} {Theory} of {Crime} and
  {Delinquency}.
\newblock {\em Criminology}, 30(1):47--88.

\bibitem[Agnew, 1999]{AgnewGeneralStrainTheory1999}
Agnew, R. (1999).
\newblock A {General} {Strain} {Theory} of {Community} {Differences} in {Crime}
  {Rates}.
\newblock {\em Journal of Research in Crime and Delinquency}, 36(2):123--155.

\bibitem[Arnio and Baumer, 2012]{ArnioDemographyforeclosurecrime2012}
Arnio, A.~N. and Baumer, E.~P. (2012).
\newblock Demography, foreclosure, and crime: {Assessing} spatial heterogeneity
  in contemporary models of neighborhood crime rates.
\newblock {\em Demographic Research}, 26:449--486.
\newblock Publisher: Max-Planck-Gesellschaft zur Foerderung der Wissenschaften.

\bibitem[Ashby, 2020]{AshbyInitialevidencerelationship2020}
Ashby, M. P.~J. (2020).
\newblock Initial evidence on the relationship between the coronavirus pandemic
  and crime in the {United} {States}.
\newblock {\em Crime Science}, 9(1):6.

\bibitem[Barari et~al., 2020]{BarariEvaluatingCOVID19Public2020}
Barari, S., Caria, S., Davola, A., Falco, P., Fetzer, T., Fiorin, S., Hensel,
  L., Ivchenko, A., Jachimowicz, J., King, G., Kraft-Todd, G., Ledda, A.,
  MacLennan, M., Mutoi, L., Pagani, C., Reutskaja, E., and Slepoi, F.~R.
  (2020).
\newblock Evaluating {COVID}-19 {Public} {Health} {Messaging} in {Italy}:
  {Self}-{Reported} {Compliance} and {Growing} {Mental} {Health} {Concerns}.
\newblock {\em medRxiv}, page 2020.03.27.20042820.
\newblock Publisher: Cold Spring Harbor Laboratory Press.

\bibitem[Bazargan, 1994]{BazarganEffectsHealthEnvironmental1994}
Bazargan, M. (1994).
\newblock The {Effects} of {Health}, {Environmental}, and
  {Socio}-{Psychological} {Variables} on {Fear} of {Crime} and its
  {Consequences} among {Urban} {Black} {Elderly} {Individuals}.
\newblock {\em The International Journal of Aging and Human Development},
  38(2):99--115.
\newblock Publisher: SAGE Publications Inc.

\bibitem[Blau and Blau, 1982]{BlauCostInequalityMetropolitan1982}
Blau, J.~R. and Blau, P.~M. (1982).
\newblock The {Cost} of {Inequality}: {Metropolitan} {Structure} and {Violent}
  {Crime}.
\newblock {\em American Sociological Review}, 47(1):114--129.
\newblock Publisher: [American Sociological Association, Sage Publications,
  Inc.].

\bibitem[Booth et~al., 1976]{BoothCrowdingUrbancrime1976}
Booth, A., Welch, S., and Johnson, D.~R. (1976).
\newblock Crowding and {Urban} crime rates.
\newblock {\em Urban Affairs Review}, 11(3):291--308.
\newblock Publisher: SAGE Publications Inc.

\bibitem[Brantingham and Brantingham, 1984]{BrantinghamPatternsCrime1984}
Brantingham, P.~J. and Brantingham, P.~L. (1984).
\newblock {\em Patterns in {Crime}}.
\newblock Macmillan.

\bibitem[Brodersen et~al., 2015]{BrodersenInferringcausalimpact2015}
Brodersen, K.~H., Gallusser, F., Koehler, J., Remy, N., and Scott, S.~L.
  (2015).
\newblock Inferring causal impact using {Bayesian} structural time-series
  models.
\newblock {\em Annals of Applied Statistics}, 9:247--274.

\bibitem[Burraston et~al., 2018]{BurrastonRelativeAbsoluteDeprivation2018}
Burraston, B., McCutcheon, J.~C., and Watts, S.~J. (2018).
\newblock Relative and {Absolute} {Deprivation}’s {Relationship} {With}
  {Violent} {Crime} in the {United} {States}: {Testing} an {Interaction}
  {Effect} {Between} {Income} {Inequality} and {Disadvantage}.
\newblock {\em Crime \& Delinquency}, 64(4):542--560.
\newblock Publisher: SAGE Publications Inc.

\bibitem[Campedelli et~al., 2020]{CampedelliExploringEffect2019nCoV2020}
Campedelli, G.~M., Aziani, A., and Favarin, S. (2020).
\newblock Exploring the {Effect} of 2019-{nCoV} {Containment} {Policies} on
  {Crime}: {The} {Case} of {Los} {Angeles}.
\newblock {\em arXiv:2003.11021 [econ, q-fin, stat]}.
\newblock arXiv: 2003.11021.

\bibitem[Chandola, 2001]{Chandolafearcrimearea2001}
Chandola, T. (2001).
\newblock The fear of crime and area differences in health.
\newblock {\em Health \& Place}, 7(2):105--116.

\bibitem[{Chicago Department of Public Health},
  2016]{ChicagoDepartmentofPublicHealthChicagoOverallHealth2016}
{Chicago Department of Public Health} (2016).
\newblock Chicago {Overall} {Health} {Status} - {Community} {Areas} (2014-2016
  {Healthy} {Chicago} {Survey}).
\newblock Dataset.

\bibitem[{Chicago Department of Public Health},
  2018]{ChicagoDepartmentofPublicHealthChicagoNeighborhoodSafety2018}
{Chicago Department of Public Health} (2018).
\newblock Chicago {Neighborhood} {Safety} - {Community} {Areas} (2016-2018
  {Healthy} {Chicago} {Survey}).
\newblock Dataset.

\bibitem[{Chicago Department of Public Health},
  2020]{IllinoisNationalElectronicDiseaseSurveillanceSystemChicagoCOVID19Cases2020}
{Chicago Department of Public Health} (2020).
\newblock Chicago {COVID}-19 {Cases}, {Tests}, and {Deaths} by {ZIP} {Code}.
\newblock Dataset.

\bibitem[{Chicago Police Department},
  2016]{ChicagoPoliceDepartmentPoliceStations2016}
{Chicago Police Department} (2016).
\newblock Police {Stations}.
\newblock Dataset.

\bibitem[{Chicago Police Department},
  2020]{ChicagoPoliceDepartmentCrimes2001Present2020}
{Chicago Police Department} (2020).
\newblock Crimes - 2001 to {Present}.
\newblock Dataset.

\bibitem[Cohen and Felson, 1979]{CohenSocialChangeCrime1979}
Cohen, L.~E. and Felson, M. (1979).
\newblock Social {Change} and {Crime} {Rate} {Trends}: {A} {Routine} {Activity}
  {Approach}.
\newblock {\em American Sociological Review}, 44(4):588--608.
\newblock Publisher: [American Sociological Association, Sage Publications,
  Inc.].

\bibitem[Cohen and Land, 1987]{CohenAgeStructureCrime1987}
Cohen, L.~E. and Land, K.~C. (1987).
\newblock Age {Structure} and {Crime}: {Symmetry} {Versus} {Asymmetry} and the
  {Projection} of {Crime} {Rates} {Through} the 1990s.
\newblock {\em American Sociological Review}, 52(2):170--183.
\newblock Publisher: [American Sociological Association, Sage Publications,
  Inc.].

\bibitem[Coveney, 2015]{CoveneyFIRTHLOGITStatamodule2015}
Coveney, J. (2015).
\newblock {FIRTHLOGIT}: {Stata} module to calculate bias reduction in logistic
  regression.
\newblock Language: en Publication Title: Statistical Software Components.

\bibitem[Damm and Dustmann, 2014]{DammDoesGrowingHigh2014}
Damm, A.~P. and Dustmann, C. (2014).
\newblock Does {Growing} {Up} in a {High} {Crime} {Neighborhood} {Affect}
  {Youth} {Criminal} {Behavior}?
\newblock {\em American Economic Review}, 104(6):1806--1832.

\bibitem[Di~Tella and Schargrodsky, 2004]{DiTellaPoliceReduceCrime2004}
Di~Tella, R. and Schargrodsky, E. (2004).
\newblock Do {Police} {Reduce} {Crime}? {Estimates} {Using} the {Allocation} of
  {Police} {Forces} after a {Terrorist} {Attack}.
\newblock {\em The American Economic Review}, 94(1):115--133.
\newblock Publisher: American Economic Association.

\bibitem[Dumas et~al., 2020]{DumasWhatDoesAdolescent2020}
Dumas, T.~M., Ellis, W., and Litt, D.~M. (2020).
\newblock What {Does} {Adolescent} {Substance} {Use} {Look} {Like} {During} the
  {COVID}-19 {Pandemic}? {Examining} {Changes} in {Frequency}, {Social}
  {Contexts}, and {Pandemic}-{Related} {Predictors}.
\newblock {\em Journal of Adolescent Health}, 67(3):354--361.

\bibitem[Firth, 1993]{FirthBiasReductionMaximum1993}
Firth, D. (1993).
\newblock Bias {Reduction} of {Maximum} {Likelihood} {Estimates}.
\newblock {\em Biometrika}, 80(1):27--38.
\newblock Publisher: [Oxford University Press, Biometrika Trust].

\bibitem[Flango and Sherbenou, 1976]{FlangoPovertyUrbanizationCrime1976}
Flango, V.~E. and Sherbenou, E.~L. (1976).
\newblock Poverty, {Urbanization}, and {Crime}.
\newblock {\em Criminology}, 14(3):331--346.
\newblock \_eprint:
  https://onlinelibrary.wiley.com/doi/pdf/10.1111/j.1745-9125.1976.tb00027.x.

\bibitem[Freedman, 1975]{FreedmanCrowdingbehavior1975}
Freedman, J.~L. (1975).
\newblock {\em Crowding and behavior}.
\newblock Crowding and behavior. W. H. Freedman, Oxford, England.
\newblock Pages: viii, 177.

\bibitem[Freeman et~al., 1996]{FreemanSpatialConcentrationCrime1996b}
Freeman, S., Grogger, J., and Sonstelie, J. (1996).
\newblock The {Spatial} {Concentration} of {Crime}.
\newblock {\em Journal of Urban Economics}, 40(2):216--231.

\bibitem[Gerell et~al., 2020]{GerellMinorcovid19association2020}
Gerell, M., kardell, j., and Kindgren, J. (2020).
\newblock Minor covid-19 association with crime in {Sweden}, a five week follow
  up.
\newblock preprint, SocArXiv.

\bibitem[Gibbs and Erickson, 1976]{GibbsCrimeRatesAmerican1976}
Gibbs, J.~P. and Erickson, M.~L. (1976).
\newblock Crime {Rates} of {American} {Cities} in an {Ecological} {Context}.
\newblock {\em American Journal of Sociology}, 82(3):605--620.
\newblock Publisher: The University of Chicago Press.

\bibitem[Glaeser and Sacerdote, 1999]{GlaeserWhyThereMore1999}
Glaeser, E.~L. and Sacerdote, B. (1999).
\newblock Why is {There} {More} {Crime} in {Cities}?
\newblock {\em Journal of Political Economy}, 107(S6):S225--S258.
\newblock Publisher: The University of Chicago Press.

\bibitem[Gowen et~al., 2020]{GowenYoungpeopleurged}
Gowen, A., Hernández, A.~R., and Rozsa, L. (2020).
\newblock Young people urged to take virus more seriously as pandemic worsens
  in {U}.{S}.
\newblock {\em Washington Post}.

\bibitem[Guerette and Bowers, 2009]{GueretteAssessingExtentCrime2009}
Guerette, R.~T. and Bowers, K.~J. (2009).
\newblock Assessing the {Extent} of {Crime} {Displacement} and {Diffusion} of
  {Benefits}: {A} {Review} of {Situational} {Crime} {Prevention}
  {Evaluations}*.
\newblock {\em Criminology}, 47(4):1331--1368.
\newblock \_eprint:
  https://onlinelibrary.wiley.com/doi/pdf/10.1111/j.1745-9125.2009.00177.x.

\bibitem[Halford et~al., 2020]{HalfordCoronaviruscrimeSocial2020}
Halford, E., Dixon, A., Farrell, G., Malleson, N., and Tilley, N. (2020).
\newblock Coronavirus and crime: {Social} distancing, lockdown and the mobility
  elasticity of crime.
\newblock preprint, SocArXiv.

\bibitem[Harper et~al., 2020]{HarperFunctionalFearPredicts2020}
Harper, C.~A., Satchell, L.~P., Fido, D., and Latzman, R.~D. (2020).
\newblock Functional {Fear} {Predicts} {Public} {Health} {Compliance} in the
  {COVID}-19 {Pandemic}.
\newblock {\em International Journal of Mental Health and Addiction}, pages
  1--14.

\bibitem[{Heartland Alliance Data},
  2018a]{HeartlandAllianceDataChicagoPeopleAged2018}
{Heartland Alliance Data} (2018a).
\newblock Chicago: {People} {Aged} 17 or less - {Community} {Areas}.
\newblock Dataset.

\bibitem[{Heartland Alliance Data},
  2018b]{HeartlandAllianceDataChicagoPeopleAged2018a}
{Heartland Alliance Data} (2018b).
\newblock Chicago: {People} {Aged} 65 or more - {Community} {Areas}.
\newblock Dataset.

\bibitem[Heinze and Schemper, 2002]{Heinzesolutionproblemseparation2002}
Heinze, G. and Schemper, M. (2002).
\newblock A solution to the problem of separation in logistic regression.
\newblock {\em Statistics in Medicine}, 21(16):2409--2419.
\newblock \_eprint: https://onlinelibrary.wiley.com/doi/pdf/10.1002/sim.1047.

\bibitem[Hipp, 2010]{HippResidentPerceptionsCrime2010}
Hipp, J.~R. (2010).
\newblock Resident {Perceptions} of {Crime} and {Disorder}: {How} {Much} {Is}
  “{Bias}”, and {How} {Much} {Is} {Social} {Environment} {Differences}?*.
\newblock {\em Criminology}, 48(2):475--508.
\newblock \_eprint:
  https://onlinelibrary.wiley.com/doi/pdf/10.1111/j.1745-9125.2010.00193.x.

\bibitem[Hooghe et~al., 2011]{HoogheUnemploymentInequalityPoverty2011}
Hooghe, M., Vanhoutte, B., Hardyns, W., and Bircan, T. (2011).
\newblock Unemployment, {Inequality}, {Poverty} and {Crime}: {Spatial}
  {Distribution} {Patterns} of {Criminal} {Acts} in {Belgium}, 2001–06.
\newblock {\em The British Journal of Criminology}, 51(1):1--20.
\newblock Publisher: Oxford Academic.

\bibitem[{Illinois Institute of Public Health},
  2020]{IllinoisInstituteofPublicHealthCoronavirusDisease20192020}
{Illinois Institute of Public Health} (2020).
\newblock Coronavirus {Disease} 2019 ({COVID}-19).
\newblock Technical report, IIPH.

\bibitem[Jacka et~al., 2020]{JackaDrugusepandemic2020}
Jacka, B.~P., Phipps, E., and Marshall, B.~D. (2020).
\newblock Drug use during a pandemic: {Convergent} risk of novel coronavirus
  and invasive bacterial and viral infections among people who use drugs.
\newblock {\em The International Journal on Drug Policy}.

\bibitem[Jarrell and Howsen, 1990]{JarrellTransientCrowdingCrime1990}
Jarrell, S. and Howsen, R.~M. (1990).
\newblock Transient {Crowding} and {Crime}:.
\newblock {\em American Journal of Economics and Sociology}, 49(4):483--494.
\newblock \_eprint:
  https://onlinelibrary.wiley.com/doi/pdf/10.1111/j.1536-7150.1990.tb02476.x.

\bibitem[Johnson, 2010]{Johnsonbriefhistoryanalysis2010}
Johnson, S.~D. (2010).
\newblock A brief history of the analysis of crime concentration.
\newblock {\em European Journal of Applied Mathematics}, 21(4-5):349--370.
\newblock Publisher: Cambridge University Press.

\bibitem[Larson and Garrett, 1996]{LarsonCrimeJusticeSociety1996}
Larson, C.~J. and Garrett, G.~R. (1996).
\newblock {\em Crime, {Justice}, and {Society}}.
\newblock Altamira Pr, Walnut Creek,Calif., subsequent edizione edition.

\bibitem[Leslie and Wilson, 2020]{LeslieShelteringPlaceDomestic2020}
Leslie, E. and Wilson, R. (2020).
\newblock Sheltering in {Place} and {Domestic} {Violence}: {Evidence} from
  {Calls} for {Service} during {COVID}-19.
\newblock {SSRN} {Scholarly} {Paper} ID 3600646, Social Science Research
  Network, Rochester, NY.

\bibitem[McKee and Milner, 2000]{McKeeHealthFearCrime2000}
McKee, K.~J. and Milner, C. (2000).
\newblock Health, {Fear} of {Crime} and {Psychosocial} {Functioning} in {Older}
  {People}.
\newblock {\em Journal of Health Psychology}, 5(4):473--486.
\newblock Publisher: SAGE Publications Ltd.

\bibitem[Merton, 1938]{MertonSocialStructureAnomie1938}
Merton, R.~K. (1938).
\newblock Social {Structure} and {Anomie}.
\newblock {\em American Sociological Review}, 3(5):672--682.
\newblock Publisher: [American Sociological Association, Sage Publications,
  Inc.].

\bibitem[Millson et~al., 2004]{MillsonSelfperceivedHealthCanadian2004}
Millson, P.~E., Challacombe, L., Villeneuve, P.~J., Fischer, B., Strike, C.~J.,
  Myers, T., Shore, R., Hopkins, S., Raftis, S., and Pearson, M. (2004).
\newblock Self-perceived {Health} {Among} {Canadian} {Opiate} {Users}.
\newblock {\em Canadian Journal of Public Health}, 95(2):99--103.

\bibitem[Mohler et~al., 2020]{MohlerImpactsocialdistancing2020}
Mohler, G., Bertozzi, A.~L., Carter, J., Short, M.~B., Sledge, D., Tita, G.~E.,
  Uchida, C.~D., and Brantingham, P.~J. (2020).
\newblock Impact of social distancing during {COVID}-19 pandemic on crime in
  {Los} {Angeles} and {Indianapolis}.
\newblock {\em Journal of Criminal Justice}, 68:101692.

\bibitem[Moore et~al., 2020]{MooreExperienceSocialDistancing2020}
Moore, R.~C., Lee, A., Hancock, J.~T., Halley, M., and Linos, E. (2020).
\newblock Experience with {Social} {Distancing} {Early} in the {COVID}-19
  {Pandemic} in the {United} {States}: {Implications} for {Public} {Health}
  {Messaging}.
\newblock {\em medRxiv}, page 2020.04.08.20057067.
\newblock Publisher: Cold Spring Harbor Laboratory Press.

\bibitem[Neale, 2004]{NealeMeasuringhealthScottish2004}
Neale, J. (2004).
\newblock Measuring the health of {Scottish} drug users.
\newblock {\em Health \& Social Care in the Community}, 12(3):202--211.
\newblock \_eprint:
  https://onlinelibrary.wiley.com/doi/pdf/10.1111/j.1365-2524.2004.00489.x.

\bibitem[Nemes et~al., 2009]{NemesBiasoddsratios2009}
Nemes, S., Jonasson, J.~M., Genell, A., and Steineck, G. (2009).
\newblock Bias in odds ratios by logistic regression modelling and sample size.
\newblock {\em BMC Medical Research Methodology}, 9(1):56.

\bibitem[Papachristos, 2013]{Papachristos48YearsCrime2013}
Papachristos, A.~V. (2013).
\newblock 48 {Years} of {Crime} in {Chicago}: {A} {Descriptive} {Analysis} of
  {Serious} {Crime} {Trends} from 1965 to 2013.
\newblock Technical report, Yale University.

\bibitem[Papachristos et~al., 2011]{PapachristosMoreCoffeeLess2011}
Papachristos, A.~V., Smith, C.~M., Scherer, M.~L., and Fugiero, M.~A. (2011).
\newblock More {Coffee}, {Less} {Crime}? {The} {Relationship} between
  {Gentrification} and {Neighborhood} {Crime} {Rates} in {Chicago}, 1991 to
  2005.
\newblock {\em City \& Community}, 10(3):215--240.
\newblock \_eprint:
  https://onlinelibrary.wiley.com/doi/pdf/10.1111/j.1540-6040.2011.01371.x.

\bibitem[Pare and Felson, 2014]{PareIncomeinequalitypoverty2014}
Pare, P.-P. and Felson, R. (2014).
\newblock Income inequality, poverty and crime across nations.
\newblock {\em The British Journal of Sociology}, 65(3):434--458.
\newblock \_eprint:
  https://onlinelibrary.wiley.com/doi/pdf/10.1111/1468-4446.12083.

\bibitem[Patterson, 1991]{PattersonPovertyIncomeInequality1991}
Patterson, E.~B. (1991).
\newblock Poverty, {Income} {Inequality}, and {Community} {Crime} {Rates}.
\newblock {\em Criminology}, 29(4):755--776.
\newblock \_eprint:
  https://onlinelibrary.wiley.com/doi/pdf/10.1111/j.1745-9125.1991.tb01087.x.

\bibitem[Payne and Morgan, 2020a]{PayneCOVID19ViolentCrime2020}
Payne, J. and Morgan, A. (2020a).
\newblock {COVID}-19 and {Violent} {Crime}: {A} comparison of recorded offence
  rates and dynamic forecasts ({ARIMA}) for {March} 2020 in {Queensland},
  {Australia}.
\newblock Technical Report g4kh7, Center for Open Science.
\newblock Publication Title: SocArXiv.

\bibitem[Payne and Morgan, 2020b]{PaynePropertyCrimeCOVID192020}
Payne, J. and Morgan, A. (2020b).
\newblock Property {Crime} during the {COVID}-19 {Pandemic}: {A} comparison of
  recorded offence rates and dynamic forecasts ({ARIMA}) for {March} 2020 in
  {Queensland}, {Australia}.
\newblock preprint, SocArXiv.

\bibitem[Payne et~al., ming]{PayneCOVID19SocialDistancing2020}
Payne, J., Morgan, A., and Piquero, A.~R. (Forthcoming).
\newblock {COVID19} and {Social} {Distancing} {Measures} in {Queensland}
  {Australia} are {Associated} with {Short}-{Term} {Decreases} in {Recorded}
  {Violent} {Crime}.
\newblock {\em Journal of Experimental Criminology}.

\bibitem[Pfeiffer et~al., 2005]{PfeifferMediaUseits2005}
Pfeiffer, C., Windzio, M., and Kleimann, M. (2005).
\newblock Media {Use} and its {Impacts} on {Crime} {Perception}, {Sentencing}
  {Attitudes} and {Crime} {Policy}.
\newblock {\em European Journal of Criminology}, 2(3):259--285.
\newblock Publisher: SAGE Publications.

\bibitem[Phillips, 2006]{PhillipsRelationshipAgeStructure2006}
Phillips, J.~A. (2006).
\newblock The {Relationship} between {Age} {Structure} and {Homicide} {Rates}
  in the {United} {States}, 1970 to 1999.
\newblock {\em Journal of Research in Crime and Delinquency}, 43(3):230--260.
\newblock Publisher: SAGE Publications Inc.

\bibitem[Piquero et~al., ming]{PiqueroStayingHomeStaying2020}
Piquero, A.~R., Riddell, J., Narvey, C., Reid, J.~A., and Piquero, N.~L.
  (Forthcoming).
\newblock Staying {Home}, {Staying} {Safe}? {A} {Short}-{Term} {Analysis} of
  {COVID19} on {Dallas} {Domestic} {Violence}.
\newblock {\em American Journal of Criminal Justice}.

\bibitem[{R Core Team}, 2013]{RCoreTeamLanguageEnvironmentStatistical2013}
{R Core Team} (2013).
\newblock R: {A} {Language} and {Environment} for {Statistical} {Computing}.

\bibitem[Ratcliffe and Breen, 2011]{RatcliffeCrimeDiffusionDisplacement2011}
Ratcliffe, J.~H. and Breen, C. (2011).
\newblock Crime {Diffusion} and {Displacement}: {Measuring} the {Side}
  {Effects} of {Police} {Operations}.
\newblock {\em The Professional Geographer}, 63(2):230--243.
\newblock Publisher: Routledge \_eprint:
  https://doi.org/10.1080/00330124.2010.547154.

\bibitem[Reppetto, 1976]{ReppettoCrimePreventionDisplacement1976}
Reppetto, T.~A. (1976).
\newblock Crime {Prevention} and the {Displacement} {Phenomenon}.
\newblock {\em Crime \& Delinquency}, 22(2):166--177.
\newblock Publisher: SAGE Publications Inc.

\bibitem[Sampson and Lauritsen, 1994]{SampsonViolentvictimizationoffending1994}
Sampson, R.~J. and Lauritsen, J.~L. (1994).
\newblock Violent victimization and offending: {Individual}-, situational-, and
  community-level risk factors.
\newblock In {\em Understanding and preventing violence, {Vol}. 3: {Social}
  influences}, pages 1--114. National Academy Press, Washington, DC, US.

\bibitem[Schnell et~al., 2017]{SchnellInfluenceCommunityAreas2017}
Schnell, C., Braga, A.~A., and Piza, E.~L. (2017).
\newblock The {Influence} of {Community} {Areas}, {Neighborhood} {Clusters},
  and {Street} {Segments} on the {Spatial} {Variability} of {Violent} {Crime}
  in {Chicago}.
\newblock {\em Journal of Quantitative Criminology}, 33(3):469--496.

\bibitem[Shavit and Rattner, 1988]{ShavitAgeCrimeEarly1988}
Shavit, Y. and Rattner, A. (1988).
\newblock Age, {Crime}, and the {Early} {Life} {Course}.
\newblock {\em American Journal of Sociology}, 93(6):1457--1470.
\newblock Publisher: The University of Chicago Press.

\bibitem[Shaw and McKay, 1942]{ShawJuveniledelinquencyurban1942}
Shaw, C.~R. and McKay, H.~D. (1942).
\newblock {\em Juvenile delinquency and urban areas}.
\newblock Juvenile delinquency and urban areas. University of Chicago Press,
  Chicago, IL, US.
\newblock Pages: xxxii, 451.

\bibitem[Spelman, 1993]{SpelmanAbandonedbuildingsMagnets1993}
Spelman, W. (1993).
\newblock Abandoned buildings: {Magnets} for crime?
\newblock {\em Journal of Criminal Justice}, 21(5):481--495.

\bibitem[{StataCorp}, 2015]{StataCorpStataStatisticalSoftware2015}
{StataCorp} (2015).
\newblock Stata {Statistical} {Software}: {Release} 14.

\bibitem[Sweeten et~al., 2013]{SweetenAgeExplanationCrime2013}
Sweeten, G., Piquero, A.~R., and Steinberg, L. (2013).
\newblock Age and the {Explanation} of {Crime}, {Revisited}.
\newblock {\em Journal of Youth and Adolescence}, 42(6):921--938.

\bibitem[{U.S. Census}, 2015a]{U.S.CensusChicagoCrowdedHousing2015}
{U.S. Census} (2015a).
\newblock Chicago {Crowded} {Housing} {Rate} - {Community} {Areas} (2010-2015
  {American} {Community} {Survey}).
\newblock Dataset.

\bibitem[{U.S. Census}, 2015b]{U.S.CensusChicagoVacantHousing2015}
{U.S. Census} (2015b).
\newblock Chicago {Vacant} {Housing} {Rate} - {Community} {Areas} (2010-2015
  {American} {Community} {Survey}).
\newblock Dataset.

\bibitem[{U.S. Census}, 2016]{U.S.CensusChicagoTotalPopulation2016}
{U.S. Census} (2016).
\newblock Chicago {Total} {Population} - {Community} {Areas} ( 2012-2016
  {American} {Community} {Survey}).
\newblock Dataset.

\bibitem[{U.S. Census}, 2018a]{U.S.CensusChicagoIncomeDiversity2018}
{U.S. Census} (2018a).
\newblock Chicago {Income} {Diversity} - {Community} {Areas} (2014-2018
  {American} {Community} {Survey}).
\newblock Dataset.

\bibitem[{U.S. Census}, 2018b]{U.S.CensusChicagoPovertyRate2018}
{U.S. Census} (2018b).
\newblock Chicago {Poverty} {Rate} - {Community} {Areas} (2014-2018 {American}
  {Community} {Survey}).
\newblock Dataset.

\bibitem[Vigil, 1988]{VigilStreetsocializationLocura1988}
Vigil, J. (1988).
\newblock Street socialization, {Locura} behavior, and violence among {Chicano}
  gang members.
\newblock {\em Urban Anthropology}, 12(11):45--75.

\bibitem[Weatherburn et~al., 1996]{WeatherburnCrimePerceptionReality1996}
Weatherburn, D., Matka, E., and Lind, B. (1996).
\newblock Crime {Perception} and {Reality}: {Public} {Perceptions} of the
  {Risk} of {Criminal} {Victimisation} in {Australia}.
\newblock {\em BOCSAR NSW Crime and Justice Bulletins}, page~8.
\newblock Publisher: Bureau of Crime Statistics and Research New South Wales.

\bibitem[Weisburd, 2015]{WeisburdLawCrimeConcentration2015a}
Weisburd, D. (2015).
\newblock The {Law} of {Crime} {Concentration} and the {Criminology} of
  {Place}.
\newblock {\em Criminology}, 53(2):133--157.

\bibitem[Wilkinson, 2004]{WilkinsonWhyviolencemore2004}
Wilkinson, R. (2004).
\newblock Why is violence more common where inequality is greater?
\newblock {\em Annals of the New York Academy of Sciences}, 1036:1--12.

\bibitem[Wilson and Boland, 1978]{WilsonEffectPoliceCrime1978}
Wilson, J.~Q. and Boland, B. (1978).
\newblock The {Effect} of the {Police} on {Crime}.
\newblock {\em Law \& Society Review}, 12(3):367--390.
\newblock Publisher: [Wiley, Law and Society Association].

\end{thebibliography}

\end{document}